\documentclass[12pt,times,preprint,review]{elsarticle}

\usepackage{adjustbox}
\usepackage{rotating}
\usepackage{multirow}
\usepackage{times}
\usepackage{graphicx}
\usepackage{color}
\usepackage{amsmath,bm}
\usepackage{lineno}
\usepackage{array}
\usepackage{delarray}
\usepackage{hyperref}

\begin{document}

\begin{frontmatter}


\title{Language Models for Materials Discovery and Sustainability: Progress, Challenges, and Opportunities}




\author{Zongrui Pei$^{1 \ast}$, Junqi Yin$^{2}$, Jiaxin Zhang$^{2}$ \\
\normalsize{$^{1}$New York University, New York, NY 10012, USA}\\
\normalsize{$^{2}$Oak Ridge National Laboratory, Oak Ridge, TN 37831, USA} \\
\normalsize{$^\ast$ To whom correspondence should be addressed; E-mail: peizongrui@gmail.com; zp2137@nyu.edu} 
}








\linenumbers


\begin{abstract}
Significant advancements have been made in one of the most critical branches of artificial intelligence: natural language processing (NLP). These advancements are exemplified by the remarkable success of OpenAI's GPT-3.5/4 and the recent release of GPT-4.5, which have sparked a global surge of interest akin to an NLP gold rush. In this article, we offer our perspective on the development and application of NLP and large language models (LLMs) in materials science. We begin by presenting an overview of recent advancements in NLP within the broader scientific landscape, with a particular focus on their relevance to materials science. Next, we examine how NLP can facilitate the understanding and design of novel materials and its potential integration with other methodologies. To highlight key challenges and opportunities, we delve into three specific topics: (i) the limitations of LLMs and their implications for materials science applications, (ii) the creation of a fully automated materials discovery pipeline, and (iii) the potential of GPT-like tools to synthesize existing knowledge and aid in the design of sustainable materials.
\end{abstract}


\end{frontmatter}

\tableofcontents

\newpage

\section{Introduction}

Natural language processing (NLP) algorithms and language models have been developed to mimic the language capability of humans to process texts on supercomputers \cite{manning1999foundations,indurkhya2010handbook,hirschberg2015advances,chowdhary2020natural}.
After intense training on corpora, they can analyze language-based content by correlating words and phrases.
NLP has started to reshape the way research is conducted. The new research norm differs from the traditional one because NLP can extract information from vast amounts of texts and even render some heuristic and hypothetical thought-like threads [Figure \ref{fig:figure1}].
As an important member of artificial intelligence (AI) and machine learning (ML) methods \cite{lecun2015deep,schmidhuber2015deep,krizhevsky2017imagenet}, NLP has been employed in materials science \cite{Tshitoyan2019,nie2021construction,hakimi2020time,court2020magnetic,Krenn201914370}, political science \cite{grimmer2013text,ficcadenti2019joint}, molecular biology and biomedicine \cite{krallinger2005text,zheng2019text,hirschman2012text,wang2019cross,thirunavukarasu2023large}, public health \cite{birgmeier2020amelie,Hoffmann2005,cheng2020overview,mani2020viral,wang2020cord,wang2021text,blanchard2022language}, chemistry \cite{white2023future,white2023assessment}, etc. [Table \ref{tab:methods}-\ref{tab:methods2} and Figure \ref{fig:figure1}{\bf c}] 
Various NLP methods have been proposed that use post-processed corpora as training data \cite{van2021open} [see Figure \ref{fig:figure1}{\bf a}].
One approach, known as the word-embedding method \cite{mikolov2013efficient,mikolov2013distributed,pennington2014glove,BERT2018,grand2022semantic,zhang2019biowordvec}, represents words and phrases in terms of high-dimensional vectors, and these word vectors are determined by their neighboring words in the corpora,  e.g., skip-gram \cite{mikolov2013efficient,mikolov2013distributed} and Glove \cite{pennington2014glove}. Recently, large language models (LLMs) with billions of parameters trained on trillions of tokens have shown impressive capabilities in solving general tasks \cite{birhane2023science}, such as producing complex plans, rendering scientific hypotheses, or creating poetry. Examples include OpenAI's generative language models GPT-3.5/4/4.5 \cite{chatGPT} and Meta's Llama 2/3/4 \cite{touvron2023llama}.

An essential task of NLP is extracting information for immediate usage or further processing. Exacted information includes physical quantities \cite{swain2016chemdataextractor,Court2020}, named entity recognition (NER, extracting phrases or words with special meanings known as named entities) \cite{Weston2019,kim2020inorganic,hansson2020semantic,trewartha2022quantifying}, complex industrial and manufacturing process protocols or chemical reactions \cite{mehr2020universal}, from corpora.
Synthesis information extracted from the chemistry literature can be directly used to copy established and develop novel synthesis pathways of compounds in experiments \cite{mehr2020universal}. Training language models involves a tokenization process of dividing sentences into tokens (i.e., individual words or phrases). After training with token pairs (tokens and their neighbors in the corpora), word-embedding models yield vector representations of the words.  
Assisted by the correlation of word vectors and other computational methods, NLP can yield context-similar chemical elements for materials discovery given sufficiently big corpora \cite{pei2023toward}. High- and medium-entropy alloys (HEAs and MEAs) \cite{liu2022machine} \cite{yeh2004nanostructured,cantor2004microstructural,zhang2014microstructures,miracle2017critical,george2019high,shi2021hierarchical,huang2022machine} are suitable material candidate classes to realize this design idea, due to the vast number in the compositional degrees of freedom they span and due to the available corpus size of more than 12,000 papers published hitherto \cite{cantor2020multicomponent}.

NLP and other AI methods are used in many research domains, including materials science \cite{eswarappa2023materials}, materials chemistry and sustainability \cite{raabe2019strategies,raabe2023materials}. Materials design enabled by AI and combinatorial high-throughput methods has been successfully applied in the discovery of novel materials \cite{raabe2023accelerating,rao2022machine,sandlobes2017rare,pei2020machine,feng2021high,pei2023toward,pei2015rapid,pei2019machine,pei2020relation,pei2021machine,pei2021mechanisms,li2024machine}.
Although several reviews have been available about the use of NLP in other fields, no systematic review or critical perspective has been published so far regarding language models in materials science, irrespective of the considerable success of NLP \cite{joshi1991natural,hirschberg2015advances,thessen2012applications,unsal2022learning,smith2022challenges,olivetti2020data,kononova2021opportunities,jones1994natural,li2018deep,indurkhya2010handbook}. The current review focuses on language-based key AI topics of highest relevance for materials science [see Figure \ref{fig:figure1}{\bf b}], namely, (i) the design and discovery of new materials based on word-embedding models or fine-tuned large language models, (ii) the development of knowledge-graph as a search and discovery engine for materials science and engineering, and (iii) the challenges and opportunities lurking behind the ``science of adequate materials-science-based prompt design" because, as in conventional science, the best start to finding solutions for fundamental problems lies in asking the right questions. We reveal the novel exploration opportunities language models provide that help summarize existing information and design sustainable materials.

{\it Uniqueness of this review}
This review is on the topic of language models for materials discovery and sustainability. We will discuss the progress, challenges, and opportunities covering different types of materials, with a particular focus on metallic materials. To the authors' best knowledge, we have not noticed review articles on the same topic. One major reason is that this topic has suddenly received extensive attention since 2022 when ChatGPT was released. Nonetheless, there are already sufficient papers to support such a review. The dynamic trends, progress, and challenges need a timely review and discussion in materials science. More importantly, our review is not limited to large language models; we will discuss language models of different sizes. In addition, we will check the capability of language models to solve challenging problems in materials science, such as the design of sustainable materials based on the United Nations criteria, a topic critical to our societal sustainability.

\begin{sidewaystable}
    \centering
    \caption{Natural language processing for materials science and other research domains. A few typical applications of representative algorithms are listed here.}
    \begin{tabular}{lllllccccc}
    \hline \hline
    Topics & Methods & Description & Refs \\ 
    \hline
ultrahigh-entropy alloys & skip-gram & context-similar elements for new HEA design & \cite{pei2023toward}\\
thermoelectric materials & skip-gram & new applications of existing materials & \cite{Tshitoyan2019}\\
lithium-ion battery & knowledge graph & quick-response model for information retrieval &\cite{nie2021construction} \\
perovskite-type oxide&NER and more & magnetic and superconducting phase transitions & \cite{court2020magnetic,swain2016chemdataextractor} \\
zeolite & text parsing and & automatic extraction of synthesis information & \cite{jensen2019machine} \\
 & regression model &  and trends & \\
 metal oxides &text extraction/ML &NN word labeling and dependency parse tree&\cite{kim2017materials} \\
 chemical terms & WE &identification of multiword chemical terms &\cite{huang2019representing}\\
 perovskite materials&WE/NER&prediction of precursors for perovskite materials&\cite{kim2020inorganic} \\
 solid oxide fuel cells&transformer(BERT) & NER,  classification of relation and abstract &\cite{gupta2022matscibert}\\
 biomaterials &BOW and more& annotation, analysis of biomaterials data & \cite{hakimi2020devices} \\
gold nanoparticle &TF-IDF& extraction of synthesis procedures, etc.&\cite{cruse2022text} \\
battery materials&skip-gram &detection of solar-rechargeable materials in papers &\cite{he2021prediction} \\
solar cells & SNA and more & identification of competitive research activity, etc. & \cite{porter2011tech} \\
perovskite solar cell&SHC algorithm, etc & evolution maps of the solar cell technology & \cite{li2019forecasting} \\
MOF & tokenization &discovering structure-property relationships & \cite{park2018text}\\
lithium-ion battery & NER & unrevealing researchers’ habits in reporting results & \cite{el2021can}\\
magnetic materials & NER and more & inverse design of magnetocaloric materials &\cite{court2021inverse} \\
drug discovery &semantic network & drug-drug interaction from abstracts of papers&\cite{percha2012discovery}\\

    \hline \hline
    \end{tabular}
\\ NER: named entity recognition; ML: machine learning; NN: neural network; WE: word embedding; BOW: bag of words; TF-IDF: term frequency–inverse document frequency; SNA: social network analysis; SHC: semantic hierarchical clustering; MOF: metal–organic frameworks; SVM: support vector machine; TF-IDF: term frequency–inverse document frequency; PMTC: prototype-matching text clustering.
\label{tab:methods}
\end{sidewaystable}

\begin{sidewaystable}
    \centering
    \caption{More applications of NLP for materials science and other research domains. A few typical applications of representative algorithms are listed here.}
    \begin{tabular}{lllllccccc}
    \hline \hline
    Topics & Methods & Description & Refs \\ 
    \hline
public Health &naïve Bayes/SVM & disease/location detection with n-gram/semantics &\cite{collier2008biocaster,conway2009classifying}\\
finance & PMTC &
information extraction from texts of financial reports & \cite{kloptchenko2004combining} \\
healthcare & TF-IDF & information extraction from health big data & \cite{kim2019associative} \\
quantum physics & semantic network & exploring trends in quantum-physical concepts & \cite{Krenn201914370} \\
biochemistry & transformer & multitask reaction predictions with explanation & \cite{lu2022unified,raffel2020exploring} \\
protein structure & transformer & sequence-based prediction of protein structures & \cite{jumper2021highly,baek2021accurate} \\
psychology &Transformer/ChatGPT & impact of ChatGPT on psychology& \cite{uludag2023use} \\
literature generation &Transformer/ChatGPT & OpenAI ChatGPT generated literature review for healthcare&\cite{aydin2022openai} \\
education &Transformer/ChatGPT& AI-assisted education using large language models &\cite{kung2023performance,alhawiti2014natural,susnjak2022chatgpt,clark2023investigating} \\
environmental research&Transformer/ChatGPT &retrieving information, streamlining workflows, etc. & \cite{zhu2023chatgpt}\\
    \hline \hline
    \end{tabular}
\\ The acronyms are the same as in Table 1. 
\label{tab:methods2}
\end{sidewaystable}


\section{Natural language processing and large language models}


\subsection{Basic concepts and terms}

We need to define and explain a few fundamental concepts and terms before introducing NLP methods and language models. We will briefly describe the basics using the simplest phrases below.

{\it tokenization}--Typical process in NLP to divide texts (paragraphs or sentences) into smaller units called tokens (e.g., words, subwords, phrases, and symbols). The tokens are then represented by one-hot vectors, i.e., vectors with only one non-zero component recording the token's position in the vocabulary. The token and its neighboring ones constitute pairs used to train NLP models.

{\it word-embedding models}\cite{mikolov2013efficient,mikolov2013distributed,pennington2014glove}--A class of algorithms that generate vector representations for words. Similar words are closer in the vector space, or their cosine of vectors is close to 1. Neural networks are frequently used to determine word vectors. There, the word vectors are obtained by maximizing their probability embedded in their contexts, which is defined by a parameter called window size. 

{\it skip-gram}--A two-layer word-embedding neural network with only one hidden layer \cite{mikolov2013efficient,mikolov2013distributed}. It is one of the most efficient algorithms to generate vectors from words.

{\it transformer}--A deep neural network that usually maintains separate input and output processing units (i.e., an encoder-decoder structure) with the self-attention mechanism. This mechanism differentially weighs the significance of each part of the original input and the recursively used output. They resemble recurrent neural networks; however, they process the entire input simultaneously, where the self-attention mechanism provides context for positions in the input sequence, making the approach suited for massive parallelization and less training times. 
Transformed has been used in foundational models for many NLP tasks. One key reason that led to the success of transformer-based models is that the model's capability follows the so-called scaling law \cite{scalinglaw}, i.e.,
\begin{equation}
    L \sim P^{-\alpha},
\end{equation}
where $L$ and $P$ are the model's loss (i.e., the deviation between a model's prediction and the ground truth) and the number of parameters, respectively, and $\alpha$ is a positive number (typically below 1) depending on the model's architecture and training data. This indicates that as the model's size ($P$) grows, its performance ($L$) improves accordingly. 

{\it BERT}--Abbreviation for Bidirectional Encoder Representations from Transformers, which is an encoder-only transformer-based large language model.
There are several BERT-based models, for example, BioBERT \cite{lee2020biobert}, SciBERT \cite{beltagy2019scibert}, clinicalBERT \cite{alsentzer2019publicly}, mBERT \cite{libovicky2020language}, PatentBERT \cite{lee2020patent}, FinBERT \cite{araci2019finbert}.

{\it GPT}--Abbreviation for generative pre-trained transformer, which is a decoder-only transformer-based large language model (LLM). Since GPT models are large in model parameters, training such models is prohibitively expensive. GPT models are generally exceptional in common NLP tasks and are used as foundation models for further fine-tuning. Typical examples of GPT models include OpenAI's GPT-3.5, GPT-4 and GPT-4.5 \cite{chatGPT}.

\subsection{Model architectures}
\begin{figure} 
\centering
\includegraphics[width=1.0\linewidth]{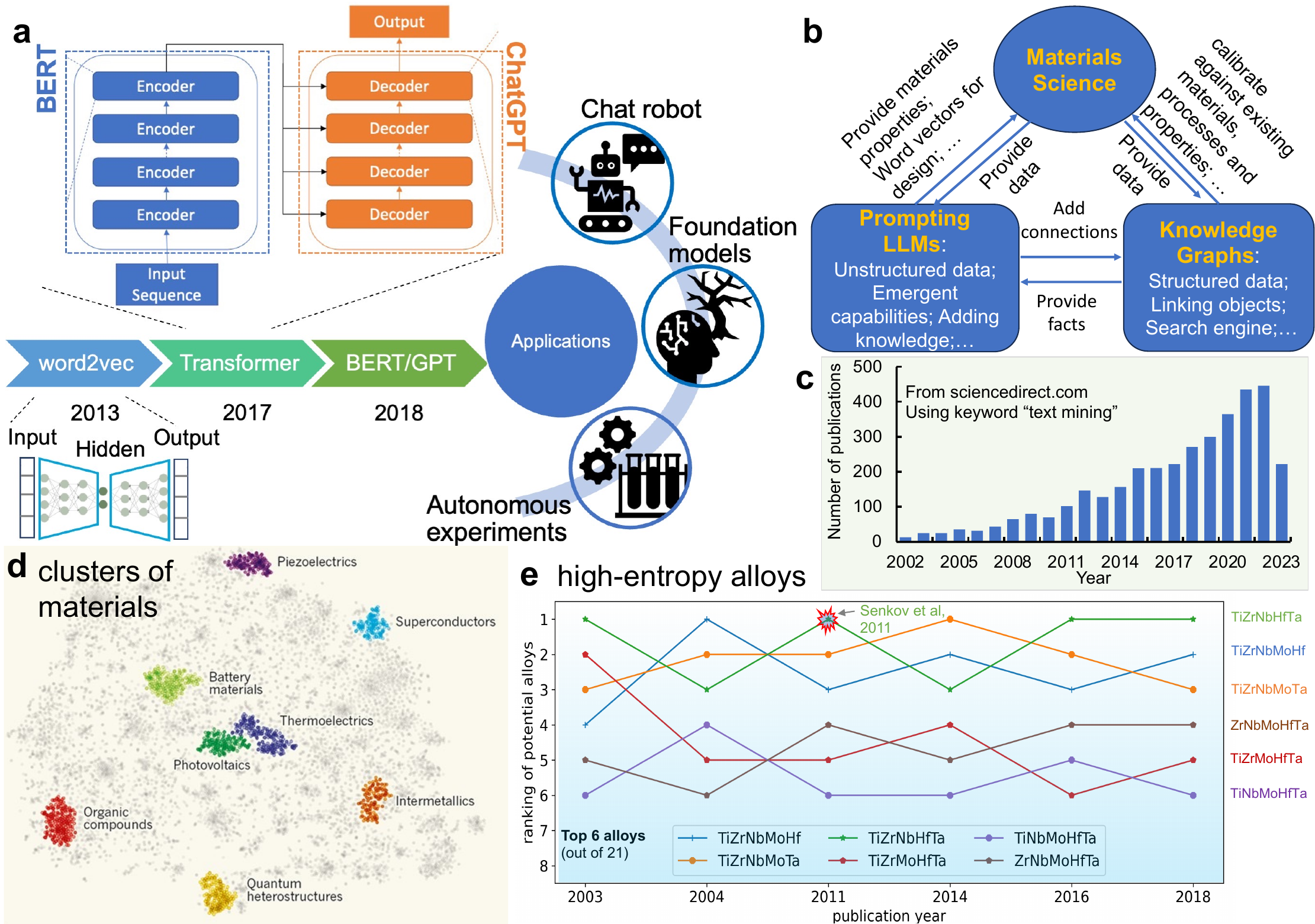}
\caption{Models and applications of natural language processing. {\bf a}, Evolution of the architectures of language models. BERT \cite{BERT2018} and GPT \cite{brown2020language} are the two main variants of transformer architecture. While BERT seems to present an upper bound \cite{hong2023diminishing} in performance, GPT-style models keep improving with data and parameter sizes, and hence become the dominant architecture for large language models. The model structures become increasingly complicated, and the model capability improves significantly with increasing model sizes and complications. For example, increasing Llama-2's model parameters from 7 billion to 70 billion can improve its accuracy by 42\% for some tasks \cite{touvron2023llama}. NLP is a multi-disciplinary research domain at the intersection of computational linguistics, computer science, and artificial intelligence. {\bf b}, Materials science interacts with prompt LLMs and knowledge-graph models. {\bf c}, Publications increase exponentially, showing the quick expansion of this field. {\bf d}, An example of scientific applications where different materials form clusters in the latent space obtained from a skip-gram model. {\bf e}, The well-known Senkov high-entropy alloy TiZrNbHfTa can be identified by a skip-gram model years earlier than its discovery. Reproduced with permission from Ref. \cite{pei2023toward,Tshitoyan2019}.} 
\label{fig:figure1}
\end{figure}

Many traditional NLP techniques, such as keyword-based search [e.g., via regular expression (RE)], word frequency-based analysis, etc., have long been applied in scientific fields \cite{superalloy}. With the introduction of word2vec and transformer architectures, NLP methods began revolutionizing how we conduct research [see Figure \ref{fig:figure1}]. We will focus on the techniques that are promising for materials research. Due to the limited space, we will not touch algorithms like seq2seq \cite{sutskever2014sequence}, LDA (latent Dirichlet allocation) \cite{pritchard2000inference,blei2003latent}, TF-IDF for text-based recommender systems \cite{anand2011mining,beel2016paper}, and non-negative matrix factorization (NMF) \cite{sra2005generalized}.  

Since the introduction of the attention mechanism in 2017, the transformer architecture has dominated the NLP field with superior performance. Three main variants, i.e., encoder-only, encoder-decoder, and decoder-only, have stemmed from the transformer family tree. From 2018 to 2019, encoder-only models such as BERT enjoyed more popularity. The decoder-only architecture has dominated the field since GPT-3, demonstrating emerging capability with unprecedented parameter size (175B). On the other hand, the number of encoder-decoder models, e.g., T5 for translation tasks, has remained about the same. While the performance of BERT-like models seems to plateau with increasing model and data sizes, as demonstrated in ScholarBERT \cite{hong2023diminishingreturnsmaskedlanguage}, the upper bound of the scaling law for GPT-like models is yet to be seen.      

While transformer-based models enjoyed many successes, the fundamental understanding of the scalable architecture is still evolving \cite{MAD}. Notably, there are shortcomings associated with the current design: (1) the computing resources required for training increase quadratically with the length of the context, which often limits its applicability despite the necessity for extensive context in many scientific applications. (2) next-token inference necessitates attention across all previous tokens, making it non-scalable and energy-inefficient.

\begin{figure} 
\centering
\includegraphics[width=0.9\linewidth]{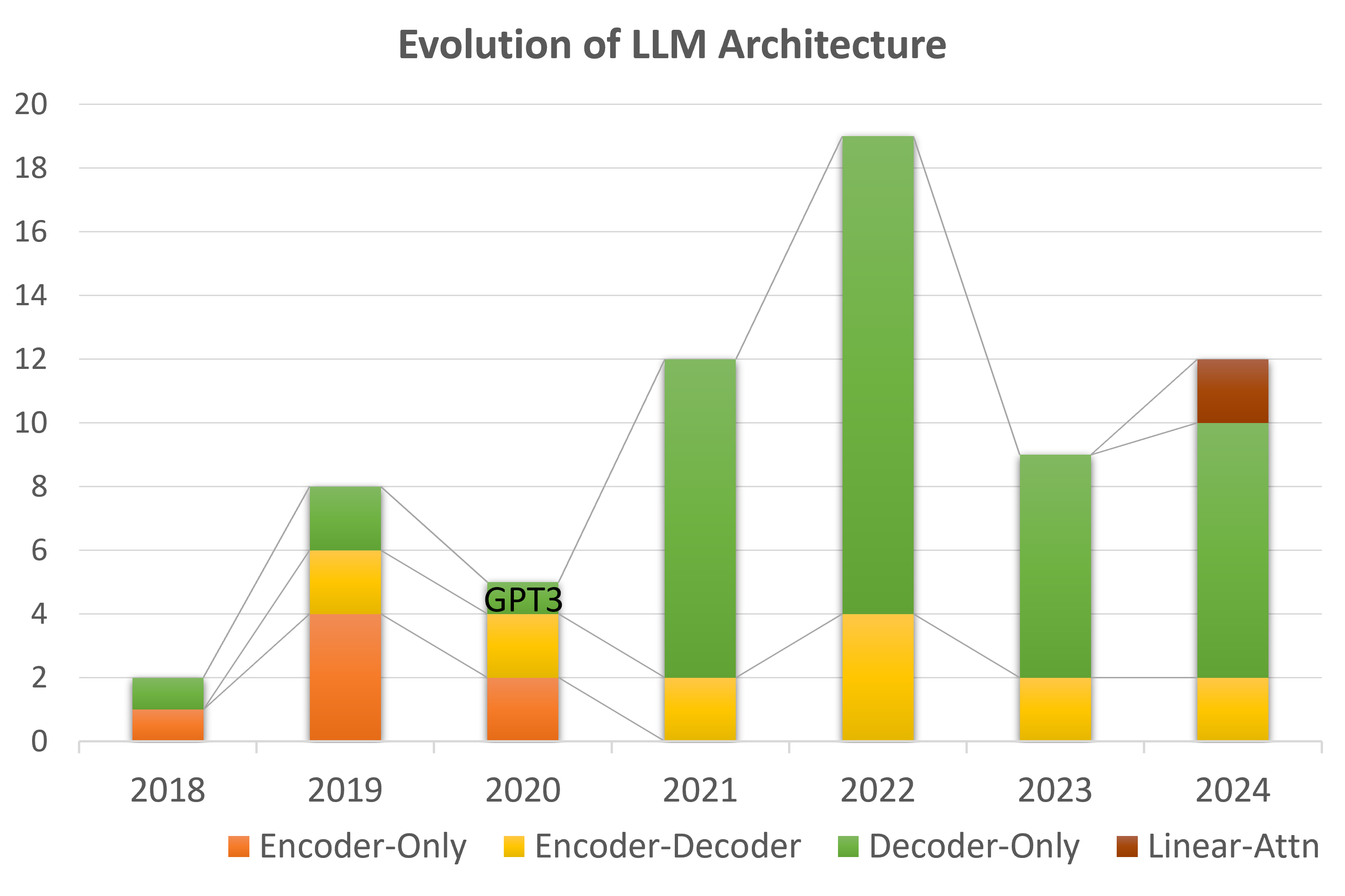}
\caption{Trends in large language model (LLM) architecture development since 2018, illustrating the distribution of Encoder-Only (e.g., BERT), Encoder-Decoder (e.g., ViT), Decoder-Only (e.g., GPT), and Linear Attention (Linear-Attn, e.g., Mamba) architectures. Notable milestones, such as the emergence of GPT-3 in 2020, are highlighted to contextualize advancements in the field.} 
\label{fig:llm-arch}
\end{figure}

The newly emerged state space model architecture \cite{mamba, rwkv}, combining the strengths of transformers and Recurrent Neural Networks, offers scalable training and inference with linear computational and memory complexities (so-called linear attention). Models such as Mamba \cite{mamba} and RWKV \cite{rwkv} have demonstrated comparable performance compared to GPT counterparts. The evolution of primary large language model (LLM) architectures is illustrated in Figure~\ref{fig:llm-arch}.

We summarize a few crucial LLMs in Table \ref{tab:LLM-features}.
The name, vendor, latest version, and key features of individual models are listed and compared. ChatGPT models are pioneering models that inspired the development of other LLMs and started the fierce competition for them. It is multimodal, supporting texts and images, designed for general chat, coding, writing, etc. Its strong plugin ecosystem renders them widely used in different platforms (mobile devices, personal computers, clouds, and supercomputers). Llama is one of the first open and the best LLMs. Given its open weights, researchers have extensively used it in customizing LLMs, particularly for fine-tuning and retrieval-augmented generation. Gemini models are known for their long context support and strong multimodal reasoning, and they have benefited from the strong Google ecosystem. Grok is closely connected with and integrated into social media. Claude models are developed with a high priority on AI safety and have strong long-context summarization. The Deepseek models are open-sourced, multimodal models that have attracted wide attention for their outstanding performance at impressively low cost. The Deepseek models have marginally close performance, compared to GPT models.

\begin{table}[]
    \centering
    \begin{tabular}{llll}
    \hline \hline
 Model name & Vendor & Latest version & Key features \\
    \hline
ChatGPT & OpenAI & GPT-4.5 & multimodal (text, image); strong reasoning \& coding    \\
Llama & Meta & Llama-4 & open weights; strong performance for fine-tuning  \\
Gemini & Google & Gemini-2.5 & long-context support; strong multimodal reasoning \\
Grok & xAI & Grok-3 & emphasizes real-time information retrieval  \\
Claude & Anthropic & Claude-3.7 & constitutional AI for safety; strong summarization/memory \\
Deepseek & Deepseek AI & DeepSeek-V3 & open weights; dedicated for multimodal and coding \\
\hline \hline
    \end{tabular}
    \caption{Representative large language models and their key features (as of April 7, 2025).}
    \label{tab:LLM-features}
\end{table}

\subsection{Foundation models}
Given the scalability of the transformer architecture, a new paradigm, the so-called foundation models, emerged in machine learning \cite{foundation-model}. The idea is to build a generic large model upon which downstream applications can be constructed. This multi-step approach is appealing since it can maximize the value of data in model training with minimal human efforts in labeling data. The abundance of unlabeled data can be leveraged to train foundation models through self-supervised learning. By capturing rich features from the data, the pre-trained model can subsequently be fine-tuned with a small set of high-quality labeled examples, enabling it to achieve strong performance on downstream tasks. The scenario applies to scientific texts as well. For example, a GPT model can be built on literature as a foundation model that encodes all scientific knowledge. For specific applications, such as the classification of alloy structures, the model can be fine-tuned on a small labeled dataset \cite{10.1145/3581784.3613215}.


Early efforts have already been made to adopt the foundation model paradigm in scientific research. For instance, in biology, an exemplar is bioGPT \cite{biogpt}, trained on a corpus of biology papers. This model has showcased its prowess in a spectrum of downstream tasks, including but not limited to relation extraction and document classification. A parallel development is evident in medicine, demonstrated by the model PubMedGPT \cite{pubmedgpt}. The performance of this model was validated in medical question and answering after leveraging its extensive training with PubMed publications. Meanwhile, climate science has also witnessed remarkable progress. Illustrated by ClimaX \cite{climatevit}, a vision transformer-based foundation model trained on simulated climate data, the capability to execute tasks like spatial downscaling and temporal prediction has been realized.

\subsection{How to build language models?}
The essential sources for training data are the scientific publishers \cite{ElsevierAPI,APS,NaturePublishing}, the rapidly growing content on pre-print and open access servers [e.g., Wikipedia], etc. 
Some particular corpora are available for dedicated domain sciences, e.g., the Matscholar NER dataset \cite{Weston2019}, the solid oxide fuel cells (SOFC) dataset \cite{friedrich-etal-2020-sofc}, and the Materials Synthesis Procedures (MSP) dataset \cite{mysore2019materials}.
To train language models, we need software and toolkits such as Scikit-Learn \cite{scikit-learn}, TensorFlow \cite{tensorflow}, PyTorch \cite{pytorch}, Natural Language Toolkit (NLTK) \cite{NLTK}, Gensim \cite{gensim}, SpaCy \cite{SpaCy}, Stanford CoreNLP \cite{CoreNLP}, AllenNLP \cite{AllenNLP}, and openNLP \cite{openNLP}, which are developed usually in C++ or Python \cite{bird2009natural}.
To speed up the training of large language models, frameworks such as DeepSpeed \cite{deepspeed} and Megatron \cite{megatron} have implemented 3D parallelisms, i.e., data, tensor, and pipeline parallelism. A model's parameters are split across multiple graphics processing units (GPUs) via tensor and pipeline parallelism, and many model replicas are trained simultaneously on different data batches via data parallelism. For an input of $D$ tokens, the total number of floating point operations needed to train a model of $P$ parameters is $6\times P\times D$. Therefore, even with an exascale supercomputer, training a one-trillion-parameter dense model will take over three months \cite{model-cost}.

Alignment is typically required to apply a language model to a specific task \cite{10.1007/s11227-024-06637-1}. This process involves continual pre-training, instruction set generation, fine-tuning, and preference tuning. (1) Continual pre-training (optional but beneficial) helps bridge the gap between the pre-trained model and the target domain. When downstream tasks demand knowledge beyond the model’s pre-training data or the outdated model needs to be updated with the latest information, continual pre-training introduces domain-specific data, enriching the model with relevant information. (2) Instruction set generation is vital for model alignment, involving the creation of high-quality instruction-response pairs. One pair usually consists of a query/question and a corresponding answer. Afterwards, automation methods and expert review are utilized to align the model performance. The reason is that methods like self-instruct can automatically streamline this process, but expert review ensures the accuracy and reliability of the instruction set. (3) The fine-tuning process uses the curated instruction set to train the model on labeled data for particular downstream tasks. Therefore, it enhances the model’s performance on domain-specific tasks. (4) Preference tuning refines the model through real-world application, employing techniques such as reinforcement learning with human feedback (RLHF). For example, collected user data and feedback are leveraged to adjust the model to offer reliable responses on customer experience. This four-step pipeline ensures the model adapts effectively to specialized domains and improves accuracy and task relevance.  


\begin{table}
\centering
\caption{NLP models for sciences.} \label{tab:models}
\begin{tabular}{lcccc}
\hline \hline
Field                     & Architecture & Model size & Data size & Ref                                \\ \hline
\multirow{4}{*}{materials science} & RE           &    $\sim$20       & 1.4M full-text         & \cite{superalloy} \\ \cline{2-5} 
                          & word2vec     &    100M      & 3$\sim$6M abstracts         & \cite{Tshitoyan2019, pei2023toward} \\ \cline{2-5} 
                          & BERT         &   110$\sim$340M           & 153K full-text          & \cite{matscibert} \\ \cline{2-5} 
                          & GPT         &   1.7$\sim$6.7B           & 27M abstract and full-text          & \cite{yin2024comparative}                          
                          \\ \hline
\multirow{2}{*}{medicine}                  & BERT         &     110$\sim$340M         &  $\sim$10B molecules         & \cite{blanchard2022language} \cite{alsentzer2019publicly} \\ \cline{2-5} & GPT & 2.7B & 35M abstracts and full-text & \cite{pubmedgpt}
\\ \hline 
\multirow{2}{*}{biology}  & BERT         &    110$\sim$340M          & $\sim$21B words         & \cite{lee2020biobert}     \\ \cline{2-5} 
                          & GPT          &    345M$\sim$1.5B          & 15M titles and abstracts         & \cite{biogpt}  \\\hline
climate                   & ViT          &   $\sim$250M         &  $\sim$PB of simulation data        & \cite{climatevit} \\ \hline \hline
\end{tabular}
\\ RE: regular expression; BERT: bidirectional encoder representations from transformers; GPT: generative pre-trained transformer; ViT: vision transformer.
\end{table}

\subsection{LLM hallucinations} 
\begin{figure} 
\centering
\includegraphics[width=1.0\linewidth]{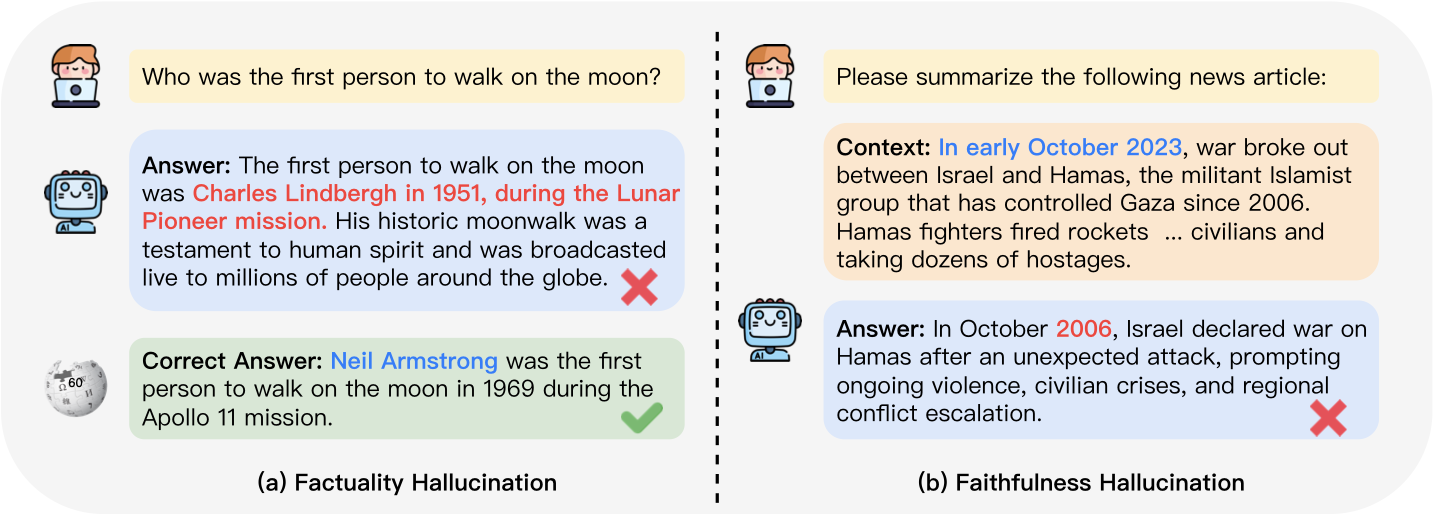}
\caption{An example of LLM hallucinations. Reproduced with permission from Ref. \cite{huang2023survey}.} 
\label{fig:hallucination_example}
\end{figure}

\begin{figure} 
\centering
\includegraphics[width=1.0\linewidth]{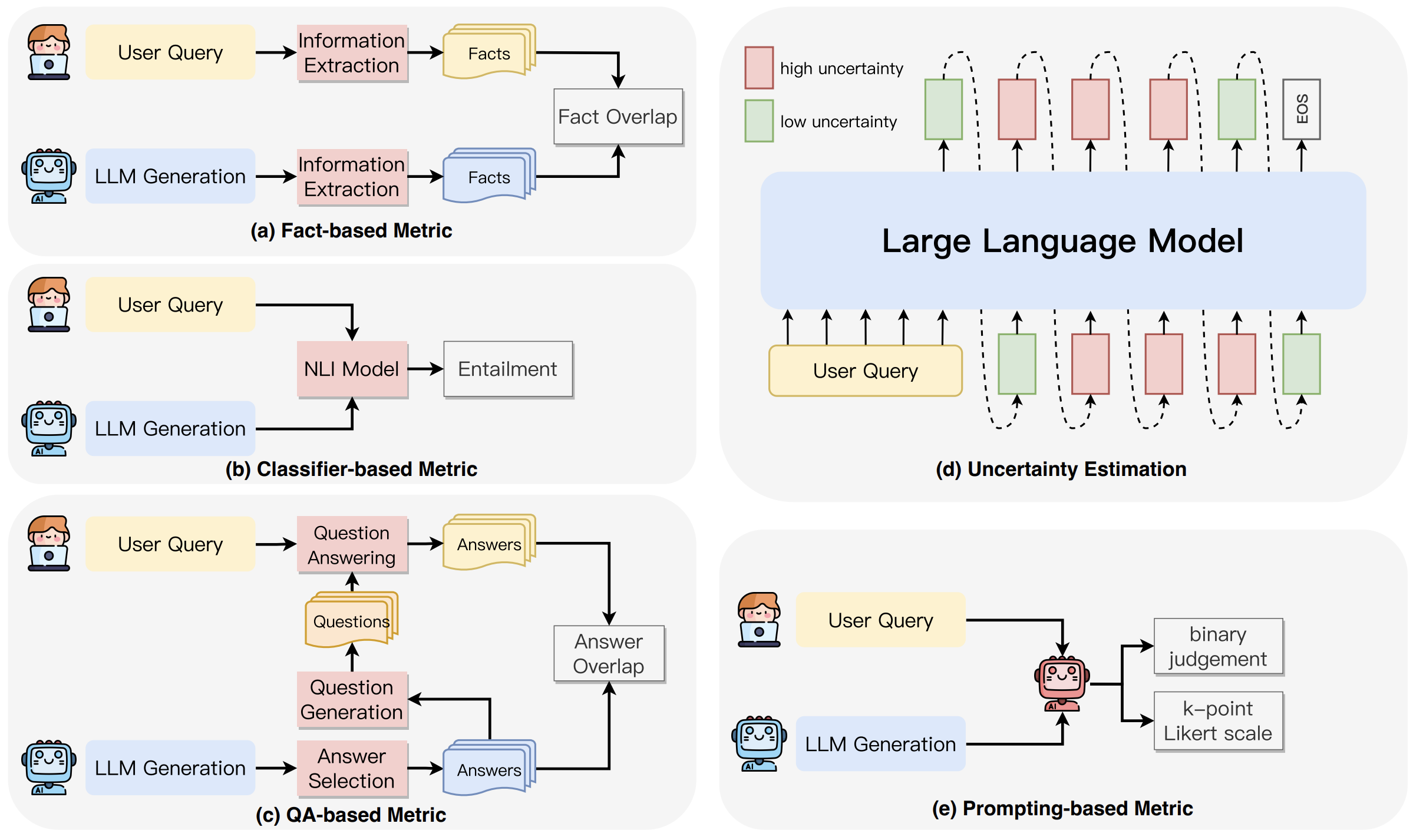}
\caption{ 
Detection methods for faithfulness hallucinations.
a) Fact-based metrics. It assesses faithfulness by measuring the factual overlap between generated content and source content.
b) Classifier-based metrics. It utilizes trained classifiers to evaluate the extent of entailment between generated content and source content.
c) QA-based metrics. It uses question-answering systems to verify the consistency of information between the source content and the generated content.
d) Uncertainty estimation. It measures faithfulness by evaluating the model’s confidence in its generated outputs.
e) Prompting-based metrics. It employs LLMs as evaluators that assess the faithfulness of generated content through specific prompting strategies. Reproduced with permission from Ref. \cite{huang2023survey}.}  
\label{fig:detection}
\end{figure}

The issue of hallucination in language models (LMs) has gained significant attention due to its negative impact on performance and the risks it introduces in various natural language processing (NLP) tasks, such as machine translation~\cite{zhou2020detecting}, summarization~\cite{cao2022hallucinated}, dialogue generation~\cite{das2023diving}, and question answering~\cite{zhang2023language,zheng2023why,dhuliawala2023chain}, see the illustrative example in Figure \ref{fig:hallucination_example}. Recent survey~\cite{ji2023survey, zhang2023siren, ye2023cognitive} and evaluation benchmarks~\cite{liu2021token,li2023halueval,yang2023new} have highlighted the importance of addressing this issue. Previous research has explored hallucination evaluation using confidence-based approaches~\cite{xiao2021hallucination, varshney2023stitch, chen2023quantifying} that require access to token-level log probability~\cite{kuhn2023semantic, cole2023selectively} or supervised tuning~\cite{agrawal2023language, li2023inference} that relies on internal states of the LM. Existing strategies for detecting hallucinations in LLMs can be categorized based on the type of hallucination: (1) factuality hallucination detection, which aims to identify factual inaccuracies in
the model’s outputs, and (2) faithfulness hallucination detection, which focuses on evaluating the faithfulness of the model’s outputs to the contextual information provided, as shown in Figure \ref{fig:detection}.

LLM hallucinations arise from various factors that span LLM's capability acquisition process. To explore hallucination origins, we categorize them into three primary aspects: (1) data, (2) training, and (3) inference. Specifically, we address data, training, and inference-related hallucinations, offering tailored solutions to tackle the specific challenges from each category. More details are referred to \cite{huang2023survey}.

Hallucinations in LLM models require a comprehensive approach to mitigate. This can be achieved through a few strategies, such as enhancing the quality of training data to ensure diversity, accuracy, and freedom from bias and implementing regularization techniques, such as dropout, weight decay, and data augmentation, which aids in preventing overfitting. In addition, adaptations to model architectures, such as employing ensemble methods or integrating attention mechanisms, can contribute to reducing hallucinations. Supplemental post-processing methods, like fact-checking, source verification, and confidence scoring, are crucial for filtering out inaccuracies. Moreover, human evaluation and feedback are essential components in refining model performance and minimizing occurrences of hallucinations.

\subsection{Retrieval-augmented generation}

Large Language Models (LLMs) demonstrate impressive capabilities but often face challenges such as hallucinations, outdated knowledge, and opaque, untraceable reasoning processes. Retrieval-augmented generation (RAG) has emerged as a promising approach that leverages knowledge from external databases, enhancing the accuracy and credibility of generated content, particularly for knowledge-intensive tasks \cite{gao2023retrieval}. This approach enables continuous knowledge updates and integrates domain-specific information with the intrinsic knowledge of LLMs.
Here, we will explore the progress of RAG paradigms and examine the tripartite foundation of RAG frameworks comprised of retrieval, generation, and augmentation techniques. We highlight cutting-edge technologies embedded within these critical components, offering a deep understanding of the advancements in RAG systems.

\begin{figure} 
\centering
\includegraphics[width=1.0\linewidth]{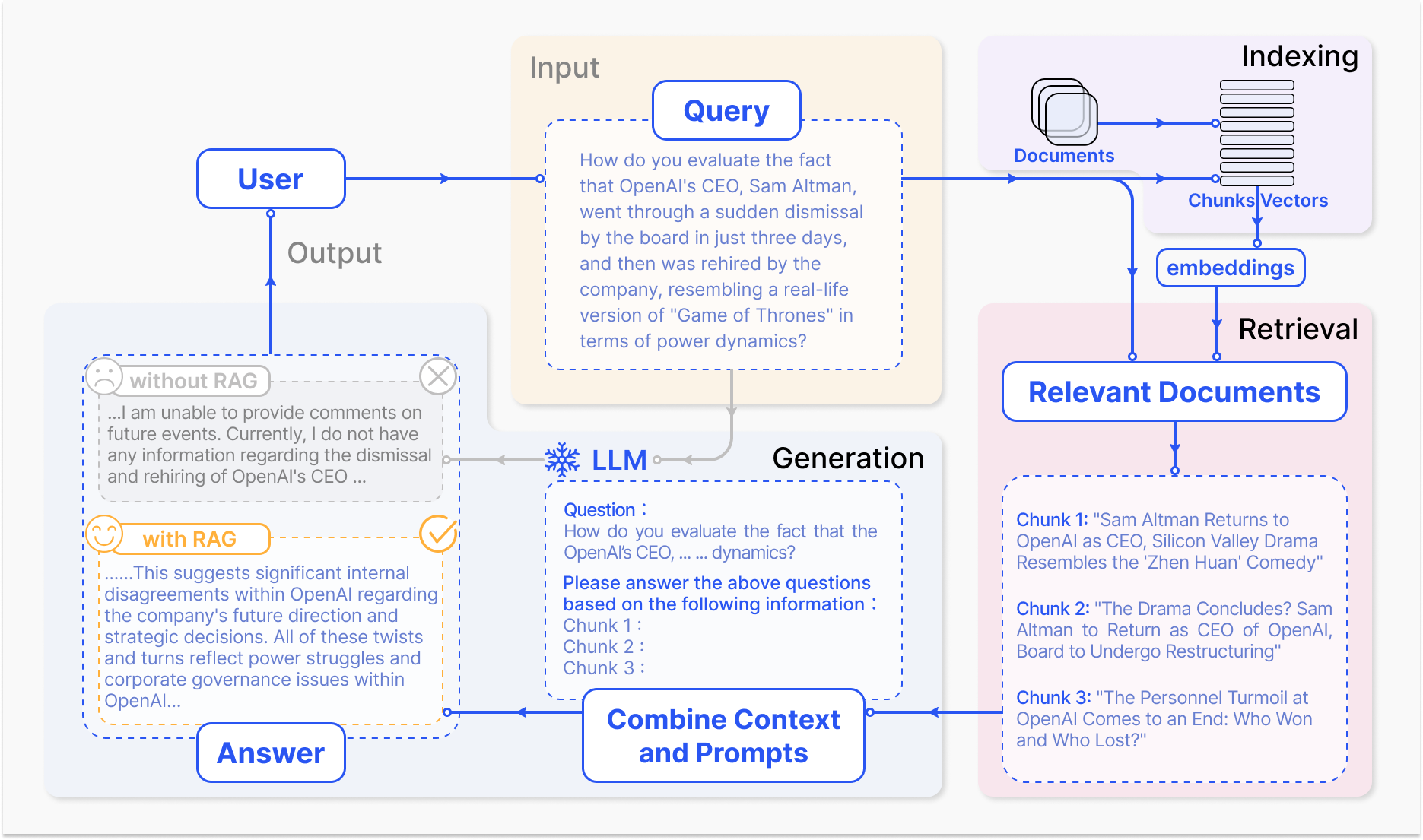}
\caption{
A representative example of the RAG process. Applying RAG to answer a question typically consists of three main steps:
(a) Indexing. Documents are segmented into chunks, encoded as vectors, and stored in a vector database. (b) Retrieval. The system retrieves the top $k$ chunks that are most relevant to the question based on semantic similarity.
(c) Generation. The original question and the retrieved chunks are taken as the input into the large language model, generating the final answer together. These steps collectively enable the RAG process to provide accurate and contextually relevant answers by integrating external, updated information. Reproduced with permission from Ref. \cite{gao2023retrieval}.} 
\label{fig:rag1}
\end{figure}

A typical application of RAG is illustrated in Figure \ref{fig:rag1}. Here, a user queries ChatGPT about a recent topic that has been widely discussed in the news. Limited by the outdated pre-training data, ChatGPT is not initially equipped with knowledge to answer this question. RAG addresses this limitation by sourcing and integrating knowledge from external databases. For this scenario, RAG retrieves relevant news articles corresponding to the user's question, enabling the LLM to generate a well-informed response based on the most current information.

\begin{figure} 
\centering
\includegraphics[width=1.0\linewidth]{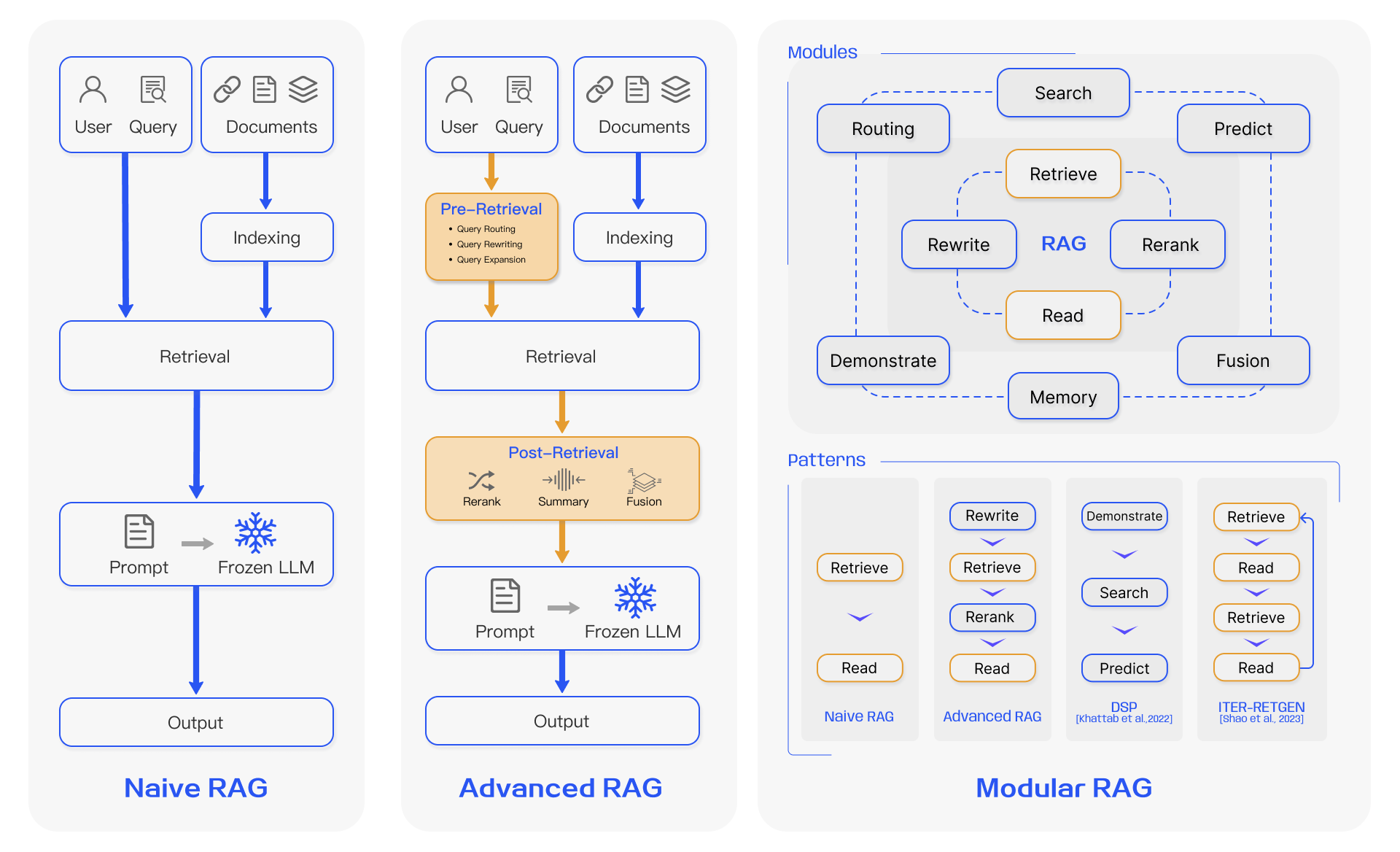}
\caption{
Comparison between the three paradigms of retrieval-augmented generation (RAG): (Left) Naive RAG primarily consists of three components, i.e., indexing, retrieval, and generation. This framework follows a straightforward process, tackling each component in sequence. (Middle) Advanced RAG builds upon the naive RAG by introducing multiple optimization strategies to enhance both the pre- and post-retrieval processes. It uses a chain-like structure to provide additional layers of refinement. (Right) Modular RAG evolves from the previous paradigm and exhibits greater flexibility. This paradigm is distinguished by its multiple specific functional modules. Unlike its predecessors, modular RAG does not strictly adhere to a sequential process. Instead, it features iterative and adaptive retrieval methods, allowing more dynamic information processing. Reproduced with permission from Ref. \cite{gao2023retrieval}.} 
\label{fig:rag2}
\end{figure}

The research paradigm of RAG is continuously evolving, and we categorize its development into three stages: naive RAG, advanced RAG, and modular RAG, as depicted in Figure \ref{fig:rag2}. Despite the cost-effectiveness and performance improvements offered by RAG methods over traditional LLMs, they also possess several limitations. Developing advanced RAG and modular RAG frameworks addresses these shortcomings inherent in the naive RAG approach.

\begin{figure} 
\centering
\includegraphics[width=1.0\linewidth]{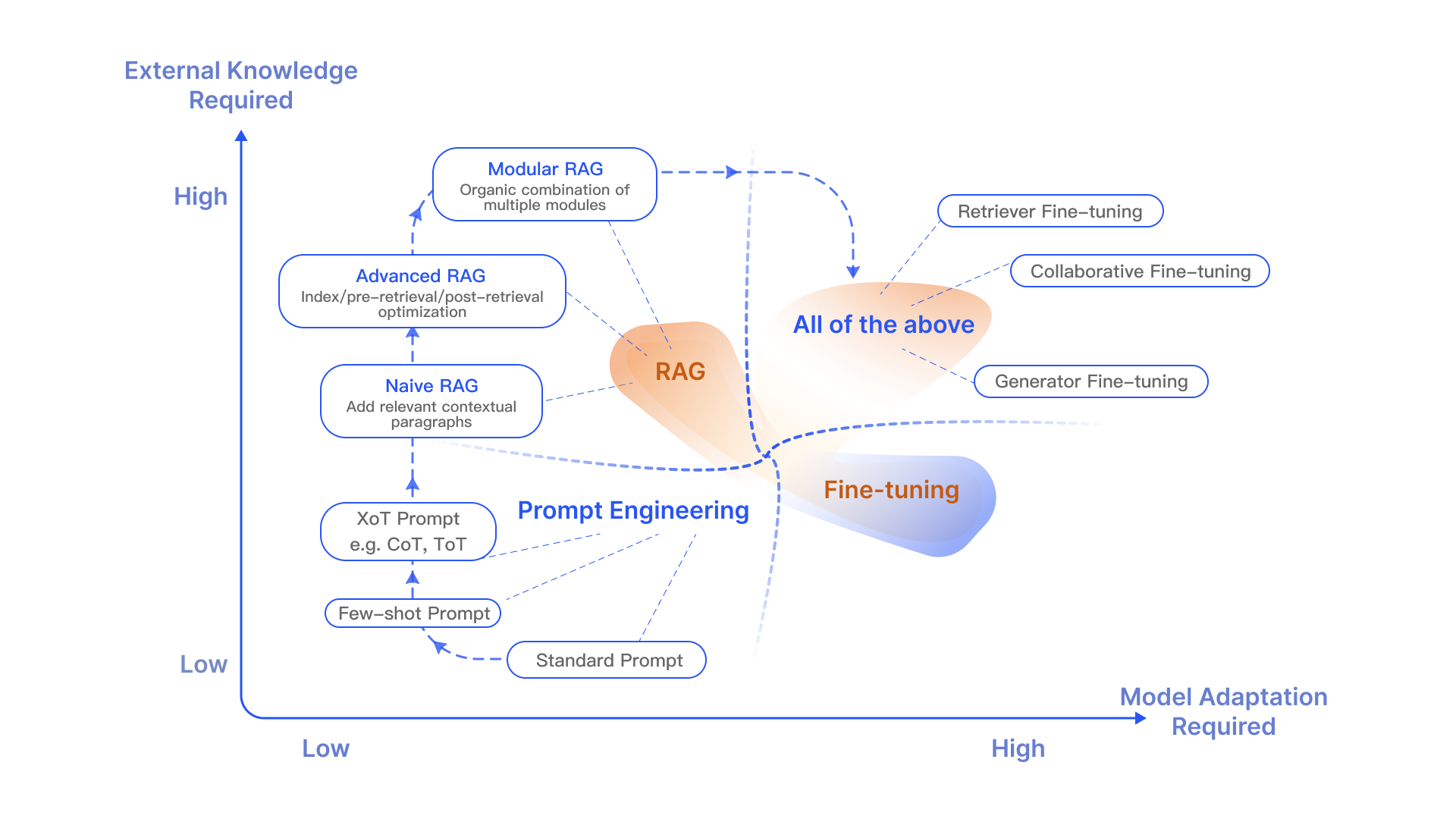}
\caption{Comparison of optimization methods like RAG, prompt engineering, and fine-tuning in terms of "External Knowledge Required" and "Model Adaptation Required." Each method has its unique advantages and disadvantages. Prompt engineering requires minimal modifications and external knowledge, focusing primarily on harnessing the existing capabilities of LLMs. Fine-tuning requires more extensive model training to tailor it for specific tasks through parameter adjustments. Initially, in naive RAG, there is a low requirement for model adjustments, utilizing external databases mainly for data augmentation. However, as the approach advances to modular RAG, there is a significant integration with fine-tuning techniques, enhancing both the model's adaptation to tasks and its ability to interact dynamically with external data sources \cite{gao2023retrieval}. Reproduced with permission from Ref. \cite{gao2023retrieval}.} 
\label{fig:rag3}
\end{figure}

RAG is often compared with fine-tuning (FT) and prompt engineering, which are among the various optimization methods for LLMs. Each method exhibits unique characteristics illustrated in Figure \ref{fig:rag3}. Here, a quadrant chart depicts the differences among these three methods across two dimensions: external knowledge requirements and model adaptation requirements.
Prompt engineering takes full advantage of the inherent capabilities of the model and requires minimal external knowledge and model tuning. This approach is conducive to leveraging the existing strengths of LLM and requires little supplementary input.
RAG can be likened to providing a model with a customized textbook for information retrieval, making it especially effective for tasks requiring precise information extraction.
In contrast, fine-tuning (FT) is analogous to a student internalizing knowledge over time. This method is particularly suitable for scenarios that require replicating specific structures, styles, or formats, adapting the model through extensive training on targeted data sets.

\subsection{Autonomous agents}

Autonomous agents have been a focus of research in academia and industry. Historically, this research direction concentrated on training agents with limited knowledge in isolated environments. This process differs from human learning and hampers the agents' ability to make decisions akin to those of humans. Nonetheless, assisted by vast amounts of external knowledge, LLMs display the potential for achieving human-level intelligence. 
Constructing agents with LLMs as their central controller is an intriguing concept. Several proof-of-concept examples have been proposed to demonstrate this potential, such as AutoGPT, GPT-Engineer, and BabyAGI. The capabilities of LLMs extend far beyond generating well-crafted text such as stories, essays, and programs; they can also be utilized as powerful general problem solvers. 

\begin{figure} 
\centering
\includegraphics[width=1.0\linewidth]{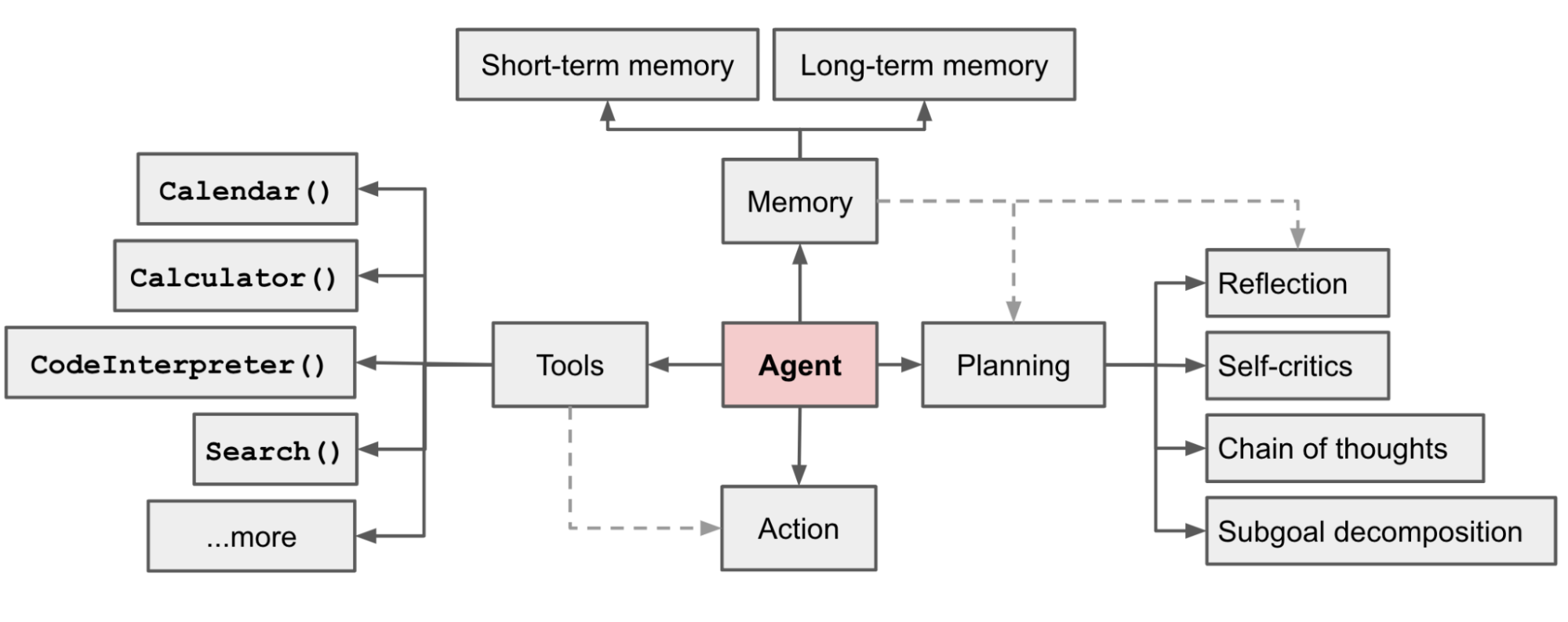}
\caption{Overview of a LLM-powered autonomous agent system. Reproduced with permission from Ref. \cite{weng2023agent}.} 
\label{fig:agent}
\end{figure}

In systems powered by LLM-based autonomous agents, the LLM functions as the cognitive core of the agent, supported by several key components detailed in \cite{weng2023agent} (see Figure \ref{fig:agent}). Here, we summarize a few of those applications below:

\begin{itemize} 
\item \textbf{Planning}. As a planner, the agent divides a large task into smaller, manageable subgoals. The agent can also engage in self-criticism and reflection on past actions, learning from mistakes and refining strategies for future steps, which enhances the overall quality of results.
\item \textbf{Memory}. The agent has the capability of recalling short-term and long-term memory. The agent can retain and recall extensive information over prolonged periods, often utilizing external vector storage and rapid retrieval mechanisms.
\item \textbf{Tool use}. The agent proficiently utilizes external APIs to access additional information not stored in the model’s weights (often difficult to modify post-training). This includes accessing up-to-date information, executing code, and reaching proprietary databases for enhanced data interaction and handling.
\end{itemize}

\begin{figure} 
\centering
\includegraphics[width=1.0\linewidth]{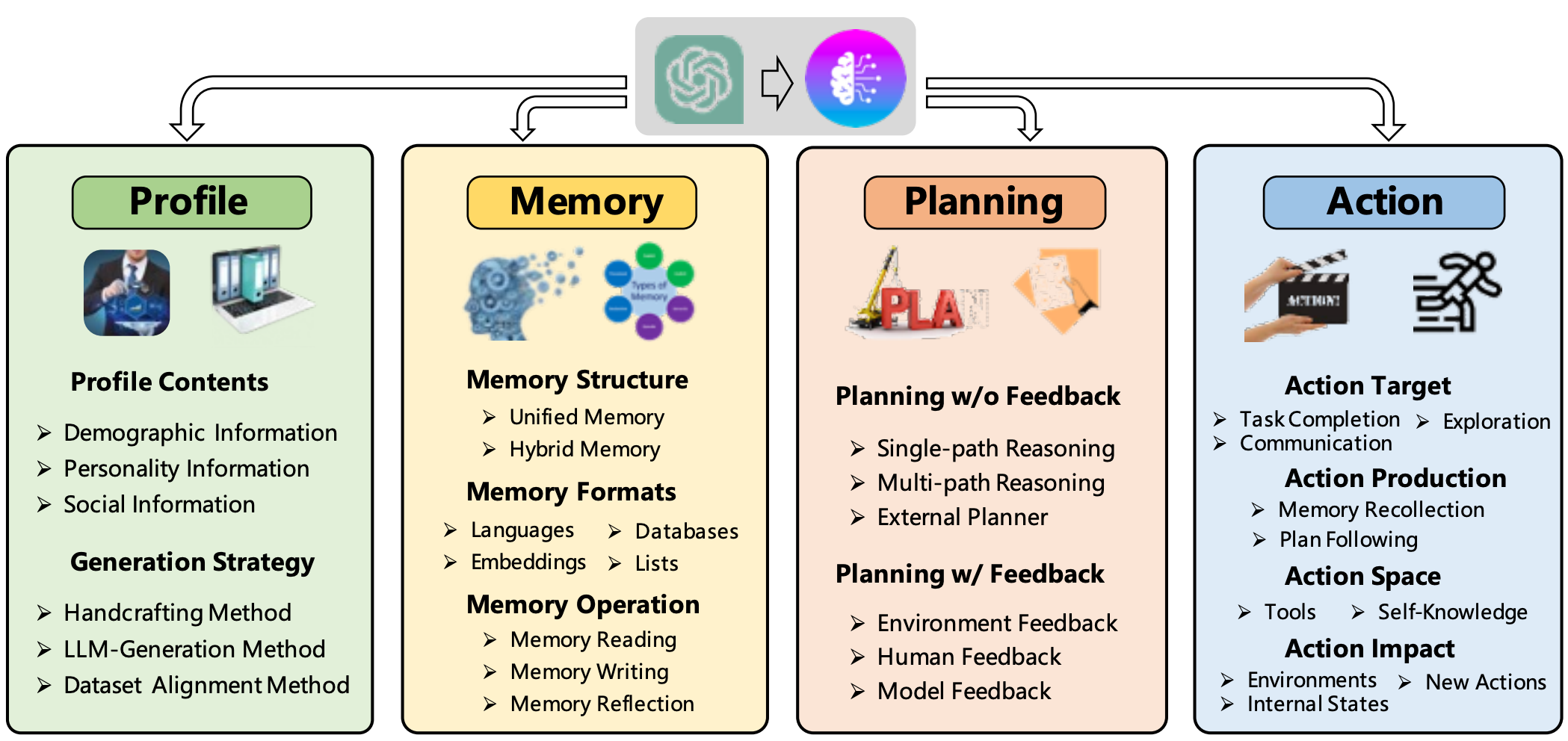}
\caption{A unified framework for the architecture design of LLM-based autonomous agent. Reproduced with permission from Ref. \cite{wang2024survey}.} 
\label{fig:agent2}
\end{figure}

Recent advancements in LLMs have shown remarkable potential in accomplishing diverse tasks through question-answering (QA). However, developing autonomous agents extends beyond the realm of QA, as these agents need to fulfil specific roles, autonomously interact with and learn from their environment, and evolve like humans. A key aspect in narrowing the gap between conventional LLMs and autonomous agents involves creating rational agent architectures to enhance LLMs. In this vein, prior research has introduced various modules to augment LLMs. The framework, depicted in Figure \ref{fig:agent2}, consists of four main components, i.e., a profiling module, a memory module, a planning module, and an action module. The profiling module determines the agent's role. In contrast, the memory and planning modules embed the agent in a dynamic environment, enabling it to retrieve past behaviors and strategize future actions. The action module transforms the agent's decisions into concrete outputs. The interplay between the profiling, memory, and planning modules substantially influences the action module. Additional information is available in Ref. \cite{wang2024survey}.

\section{What can NLP and LLMs do with materials science texts?}
\begin{figure}
    \centering
    \includegraphics[width=0.8\linewidth]{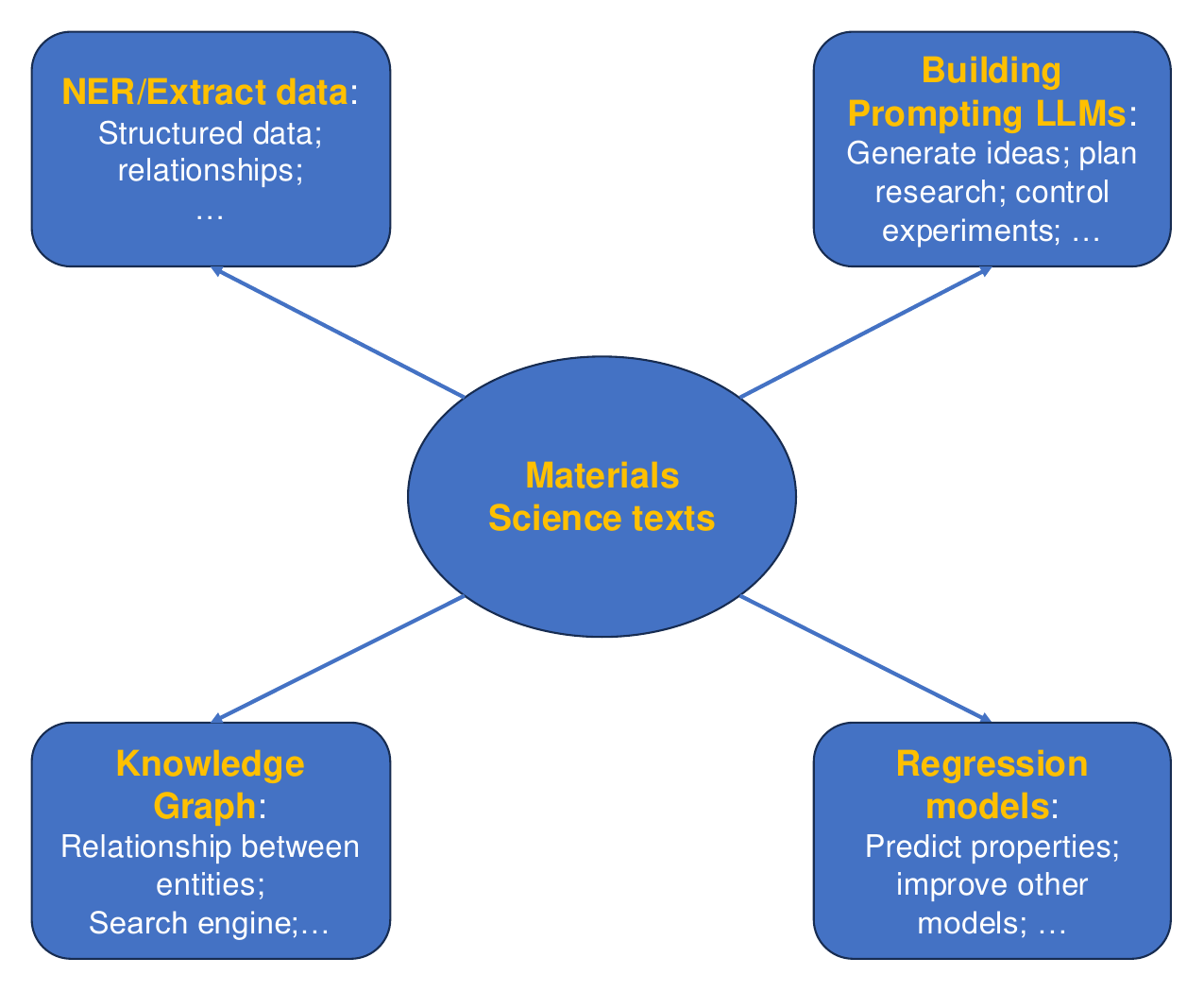}
    \caption{Natural language processing and materials science texts. We summarize four important tasks that natural language processing and language models can do with materials science texts. These tasks are not completely independent, but closely related.}
    \label{fig:what}
\end{figure}

Language models signal a paradigm shift in materials research \cite{yu2024large}, and its application is a fast-evolving research domain. Currently, we can group the applications of natural language processing to materials science texts into four essential categories (Figure \ref{fig:what}), i.e., information extraction, building prompt language models, constructing knowledge-graph models, and informing a third regression and classification models. These four groups of applications are not independent but interconnected.

These tasks do not cover all applications. For example, specific to chemistry, Guo and colleagues benchmarked eight tasks that large language models can do \cite{guo2023can}, including understanding tasks (e.g., name prediction, property prediction), reasoning (e.g., yield prediction, reaction prediction, reagents selection, retrosynthesis, text-based molecule design), and explaining (molecule captioning). These tasks largely overlap with the four functions in Figure \ref{fig:what} but are not identical. We will elaborate our discussion with more technical details and examples.

\subsection{Extract information}

Material properties or data are not always structured or not organized in tables. Instead, they usually appear in unstructured texts. Extracting information or data from peer-reviewed publications can help tackle the issue of data scarcity in training machine-learning models \cite{yu2024large}. Given the exponential trend of increased publications, this task is impractical to achieve manually. To this end, text-mining techniques, like named entity recognition (NER), are utilized. Extracting the values of the properties from texts using NER has become a critical topic. 

In addition to individual material properties, the relationship between a physical property and its measurement conditions is vital. For example, the yield stress of the same material varies largely with measuring temperature. Yield stress close to melting temperature is much smaller than measured at room temperature. We need to extract not just the value for yield stress but also the measuring temperature and store them in a pair. In addition to measuring temperature, the processing history is also essential. Then, we want to add other quantities to the pair and make a ternary relationship. This process can go to even higher orders when more factors are included. Higher-order relationships are complicated to extract, mainly due to the exponential increase in their quantities. Recently, some studies have tried to tackle this issue \cite{dagdelen2024structured}.
Dagdelen {\it et al.} proposed an approach to joint relation extraction and NER \cite{dagdelen2024structured}. This approach does not enumerate the relationships as 2-tuples or multi-tuples but stores the entities in a structured format (e.g., JSON), which significantly reduces the volumes of extracted data. They used two pre-trained large language models (GPT-3, Llama-2) to extract complex scientific knowledge after fine-tuning the models. As an example, they tested their approach on three tasks, i.e., (i) linking dopants and host materials, (ii) cataloguing metal-organic frameworks, and (iii) extracting the composition-phase-morphology-application information. The method demonstrates the high potential of large language models to extract structured data from scientific texts.

Large language models can also be used to extract data from texts.
Polak {\it et al.} used general purpose language models (i.e., GPT-3, GPT-3.5/4, BART and DeBERTaV3) to extract materials data from text, taking bulk modulus as an example \cite{polak2024flexible}. They developed a multi-step framework that includes pre-processing original texts, feeding the processed texts to language models, collecting the outputs, and finally post-processing the outputs. The extracted bulk moduli are prepared in a flexible format for future applications. They demonstrated the reliability of this method (up to 90\% precision at 96\%  recall) and the efficiency compared to manual extraction of human beings. In another separate work, the same authors proposed a method, ChatExtract, to extract materials data from research papers \cite{polak2024extracting}. Based on large language models and prompt engineering, the method consists of a few follow-up questions to make decisions in data extraction from individual sentences (discarding or continuing the process). The high-quality follow-up questions ensure the accuracy of the data.

Schilling-Wilhelmi {\it et al.} performed a comprehensive overview of data extraction from texts based on large language models \cite{schilling2024text}. They concluded this extraction process consisted of three primary steps, i.e., (i) data curation and pre-processing, (ii) interaction with large language models (fine-tuning, prompt engineering, etc), and (iii) post-processing the outputs of language models. They also discussed the principles of language models and future research directions.

\subsection{Build a prompt model}

The second pathway to utilizing materials science texts is building large language models (e.g., GPT and BERT). We have introduced the fundamental concepts and how to train a large language model in previous sections, which we do not repeat here. Given high-quality prompts, language models can generate new research ideas and plan research. It can also provide initial validation of the research plan and adjust and improve its quality. For example, it can help identify if existing research supports or goes against the plan and check if the materials intended to be designed already exist. They can also act as a control center in the future to manipulate equipment and automate experiments \cite{pei2024towards,boiko2023autonomous}.

Materials science texts can also be used to fine-tune general-purpose foundation models or update these models using the retrieval-augmented generation (RAG) technique.
For example, Chiang and colleagues developed a multimodal RAG framework, LLaMP, based on large language models \cite{chiang2024llamp}. The framework aims to mitigate hallucination and intrinsic biases of language models by retrieving external resources and databases (e.g., Materials Project) and distilling high-fidelity, high-quality materials knowledge. As proof of concept, the framework was checked for physical properties like bulk moduli, formation energies, etc. The generated properties are not simply drawn from individual databases but mixed values derived from multiple resources. The LLaMP framework also shows the capability of running atomistic simulations with empirical potentials.

\subsection{Build a knowledge graph}

The third task is utilizing materials science texts to build a knowledge graph. The knowledge graph is a model usually used in search engines like \url{google.com}. Materials scientists have a particular preference for naming materials and describing their discovery. Hence, a search engine designed for materials science is particularly welcome. Assisted by it, we can exhaust all available information on a research topic, not affected by the multiple names of a material or a physical quantity, etc. Examples will be discussed in Section \ref{sec:KG} and not repeated here.

\subsection{Provide information/features for other models}

The above three tasks are unsupervised, i.e., no labels or definite targets are involved or optimized.
In the fourth task, materials science texts are used in a supervised manner to construct a regression or classification model. The implementation details can be different and may need assistance from a foundation language model. Some researchers fine-tune the model parameters with prompt-output data and use the language models to generate outputs. Other models include one additional layer for regression tasks, i.e., actively adjust word embedding vectors with this regression layer, not passively use them from the original LLMs.
Nonetheless, these methods generate quantitative predictions using word-embedding vectors from LLMs, skip-gram models, etc. The word vectors of unstructured information (like processing history) can improve model performance \cite{sasidhar2023enhancing}. Some researchers even use LLMs (e.g., GPT, Llama) to generate quantitative answers for materials properties in regression or classification tasks after fine-tuning them \cite{jablonka2024leveraging,jacobs2024regression}. Liu {\it et al.} proposed an LLM-based model for classifying metallic glasses \cite{liu2024prompt}. These exciting proceedings push the applications of language models towards a new direction and demonstrate the enormous application opportunities.

\begin{figure}
    \centering
    \includegraphics[width=0.8\linewidth]{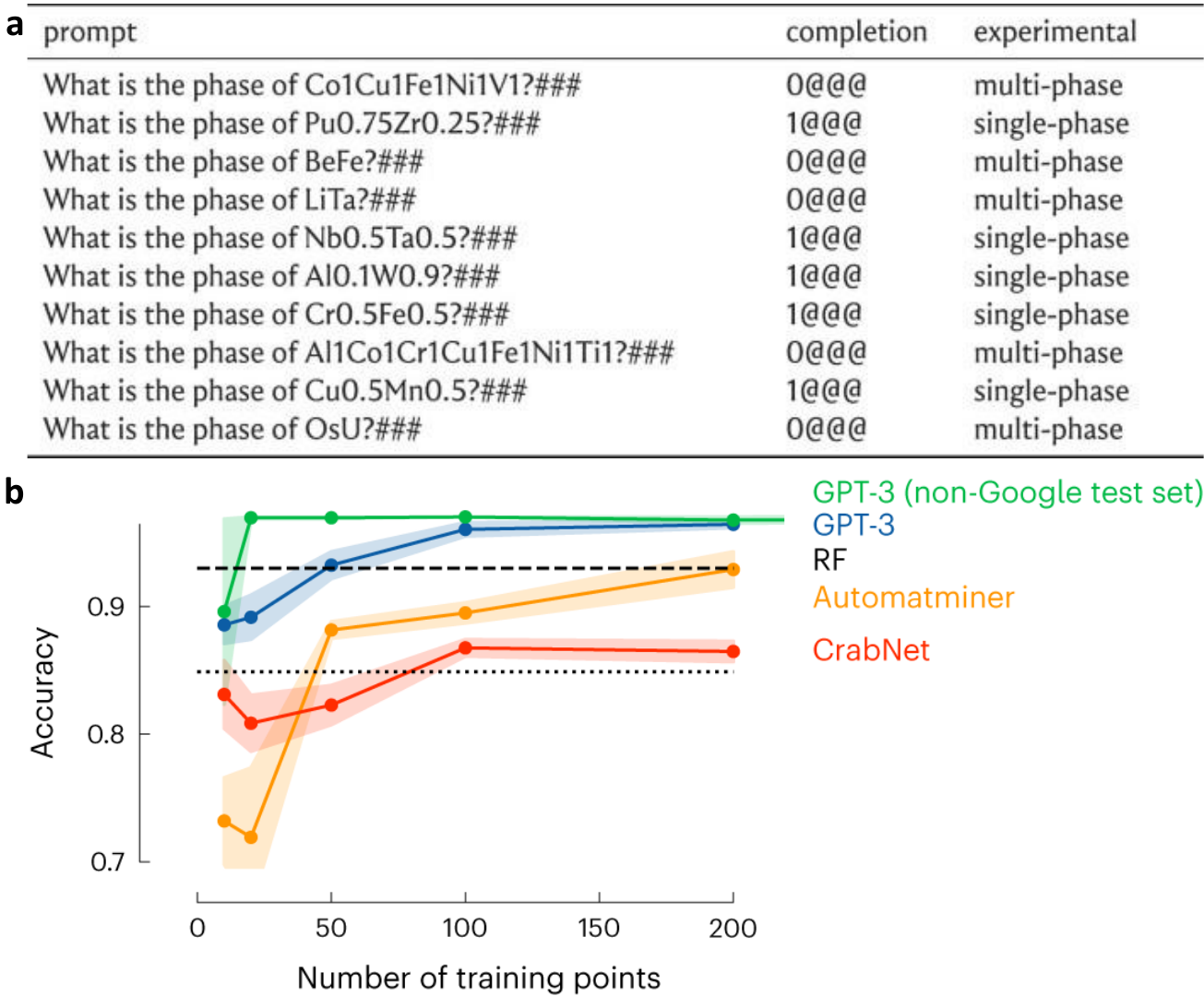}
    \caption{Fine-tuning GPT-3 for solid-solution prediction. {\bf a}, Examples of training data. They are pairs of prompts and completions. Here, the string ``\#\#\#" indicates the end of a prompt, and ``@@@" is the end of completion. Multi-phase alloys are represented by 0 and single-phase alloys by 1. The fine-tuning principle of the model is the same as GPT-3, i.e., predict the next token given a prompt or a text sequence. {\bf b}, the performance of the fine-tuned GPT-3 and other traditional machine-learning models. Reproduced with permission from Ref. \cite{jablonka2024leveraging}.}
    \label{fig:FigureLLM-Classification}
\end{figure}

Conventional machine-learning models are trained on small datasets focusing on a specific knowledge domain and need specialized expertise. LLMs offer new opportunities to solve various downstream tasks without requiring specialized expertise since they are trained on general-purpose texts that are not limited by specific knowledge. Jablonka {\it et al.} demonstrated how to leverage LLMs to predict the properties of molecules and materials as well as the yield of chemical reactions \cite{jablonka2024leveraging}. After fine-tuning, they found GPT-3 could achieve these tasks with comparable or even better performance than the conventional machine-learning models.
In one of the classification examples, the authors demonstrated the fine-tuned model could predict solid-solution formation in conventional and high-entropy alloys. Compared with three conventional models, they showed their models have superior performance with even relatively small training data (Figure \ref{fig:FigureLLM-Classification}). The reason is that the LLM already includes valuable information in the foundation model, while the other models trained on specialized datasets contain minimal information. The result confirmed the importance of the available information in the pre-trained large models.
Interestingly, in one of these tasks, the authors fine-tuned GPT-3 to predict HOMO–LUMO gaps of molecules and showed it has extrapolation capability.

\section{NLP for materials discovery and sustainability}

Materials design is an iterative process that determines suitable chemical elements, synthesis, processing, and property combinations to fulfill a specific application need. This process is a typical inverse problem which must be often started on the basis of insufficient information  \cite{sanchez2018inverse,zunger2018inverse,tarantola2006popper,yu2013inverse,pei2021machine,zhang2021efficient,fung2021inverse}. Therefore, besides the knowledge about chemical elements (e.g., the elemental properties listed in the Periodic Table of the Elements), researchers must also read the literature and extract the most relevant existing materials as a starting point to design materials. When it comes to the intrinsic properties and bonds associated with the electronic structure of the elements, quantum mechanical simulation methods are an excellent basis for materials design. However, when multiple elements and complex synthesis and processing pathways are also considered, knowledge harvesting from the written corpora can become essential. This situation is characteristic of advanced materials. Text mining can help pre-structure promising material subgroups or meaningful composition spaces. This section discusses NLP applications for various materials, including structural materials \cite{pei2023toward,pei2024towards}.

\subsection{NLP for structural materials}
Structural materials carry mechanical loads. Metals and alloys are typical structural materials that have played a critical role in the civilization of human beings. The mechanical performance of structural materials can be characterized by a wide variety of descriptors, such as elastic modulus, yield stress, toughness, hardness, abrasion resistance, and ductility in a tensile test. These features are created not only by their electronic structures but also by a complex cosmos of lattice defects that enable or suppress specific mechanical properties. In addition to the average chemical composition, these defects are also highly dependent on the processing of the materials because the processing imprints the required defect landscapes into materials. 
Therefore, designing high-performance structural materials is a nontrivial task and cannot be reduced to selecting appropriate chemical compositions alone.
ML is an effective method for developing high-performance structural materials, as shown by its success in atomistic simulations \cite{ZHANG2020108247,LIU2021110135,yin2021neural}, physical-property prediction \cite{pei2020machine, NatureCommCCAHard}, and materials design \cite{ha2021evidence, DDJohnsonML2021Nature,rao2022machine}. 
For example, they have been used to predict solid solutions, intermetallics, amorphous, and mixed phases \cite{ZHANG2020528, HUANG2019225,ISLAM2018230, TANCRET2017486,pei2020machine}, determine crystal structures (e.g., face-centered cubic, body-centered cubic, hexagonal close-packed) \cite{KAUFMANN2020178, PhysRevMaterials.3.095005,LEE2021109260}, identify important descriptors and new empirical rules \cite{ML_Phase_sensitivity, pei2019machine, pei2020machine}, and predict the mechanical properties of complex concentrated alloys \cite{NatureCommCCAHard, WEN2019109, peng2020coupling}.

\begin{figure*} 
\centering
\includegraphics[width=0.95\linewidth]{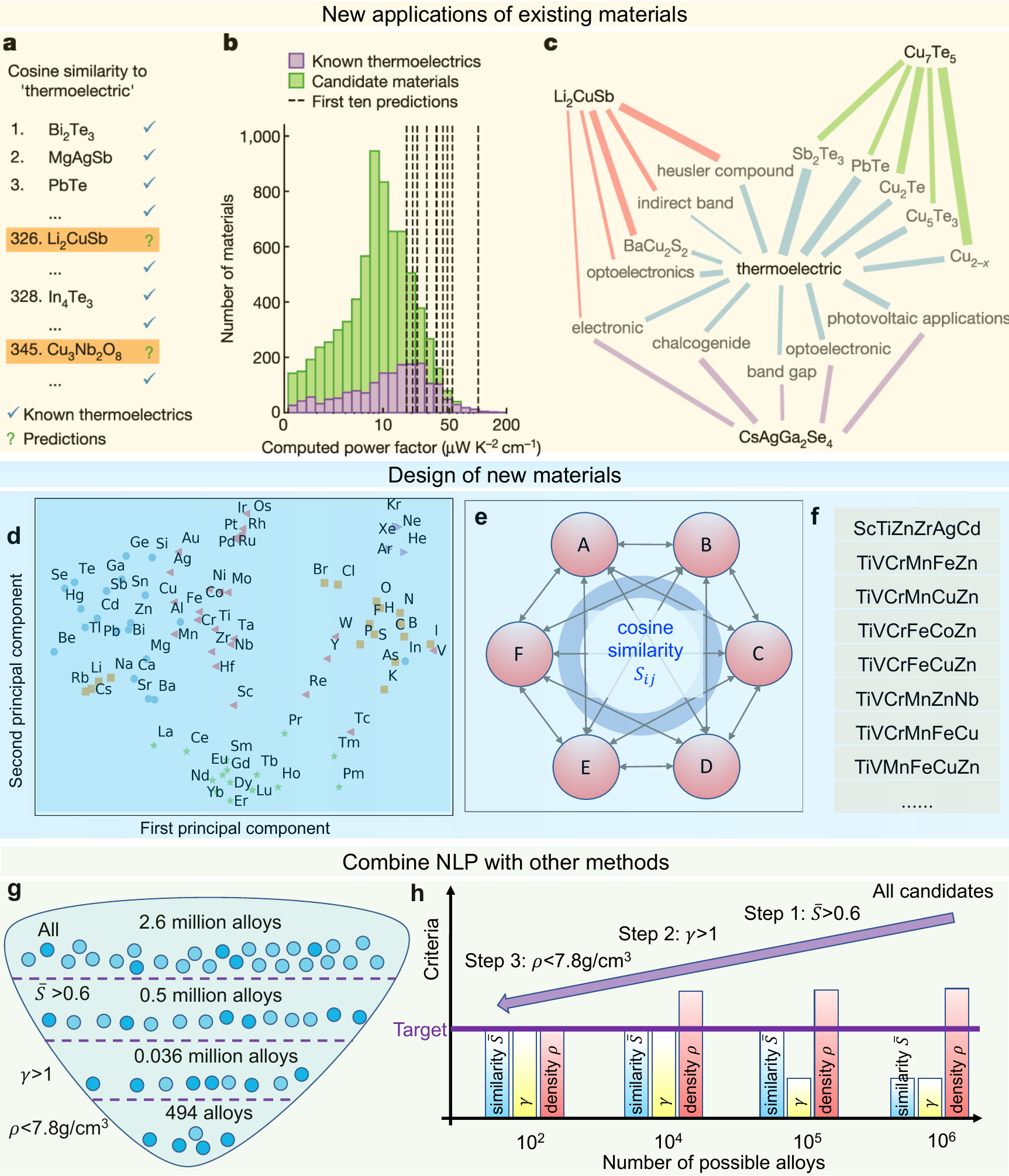}
\caption{Material-design strategies based on natural language processing. {\bf a-c}, exploring the new applications of existing materials. {\bf a}, A list of materials ordered by their cosine similarity to the word ``thermoelectric". {\bf b}, Not all the top-ranking materials are known thermoelectrics, and these unknowns are predicted to be promising candidates for synthesis. {\bf c}, The predicted unknown thermoelectrics and the word "thermoelectric" share common connections with known thermoelectrics. The line width represents the correlation's strength or the cosine similarity's magnitude {\bf d-f}, Design of new materials not mentioned in the literature. {\bf d}, The distribution of the chemical elements in the latent space from the ML model. {\bf e}, A schematic to show how to calculate the cosine similarity of an ultrahigh-entropy alloy. {\bf f}, Top candidates ranked by alloy's cosine similarity. {\bf g-h}, An ICME example to design lightweight high-entropy alloys. Reproduced with permission from Ref. \cite{Tshitoyan2019,pei2023toward}.} 
\label{fig:TM}
\end{figure*}

Recent NLP applications for material design include alloys and non-metallic materials \cite{pei2023toward,Tshitoyan2019}. To demonstrate the unique features of designing metallic alloys, we first introduce a typical work on thermoelectrics. Figure \ref{fig:TM}{\bf a}-{\bf c} shows that NLP can identify new applications of existing materials. In this study, materials that never appeared in the context of thermoelectrics were found to possess promising thermoelectric properties. More specifically, the word vectors for all materials present in the training corpora were determined by a skip-gram model. Based on the cosine similarity of these vectors, an ordered list of materials was generated [see Figure \ref{fig:TM}{\bf a}]. In this list, 1820 materials are well-known as thermoelectric materials. Nonetheless, 7663 materials in the list were not considered as thermoelectrics, but first-principles calculations confirmed their capability as thermoelectrics [see Figure \ref{fig:TM}{\bf b}]. Figure \ref{fig:TM}{\bf c} shows a possible origin underlying the predictions. The link is actually built on their common words [e.g., ``indirect band", ``optoelectronics", etc.] between the word ``thermoelectric" and the materials [e.g., Li$_2$CuSb].

The above method can only discover materials present in the corpora. This limitation hinders its broader applications in exploring novel materials like high-entropy alloys (HEAs). In one recent study \cite{pei2023toward}, NLP demonstrated the potential to design materials not reported in the literature when jointly used with other methods [Figure \ref{fig:TM}{\bf d}-{\bf f})]. Researchers introduced a concept of ``context similarity" for selecting chemical elements for HEAs based on word-embedding models that analyzed the abstracts of 6.4 million papers [see Figure \ref{fig:TM}{\bf d}-{\bf f})]. The method captures the similarity of chemical elements in the context used by scientists. It overcomes the limitations of the conventional word-embedding algorithms. The approach successfully identified the Cantor and Senkov HEAs [see Figure \ref{fig:figure1}{\bf e}]. It was then used to search for six- and seven-component lightweight HEAs and identified nearly 500 promising alloys out of 2.6 million candidates. The method thus brings an NLP-based approach to developing ultrahigh-entropy alloys and multicomponent materials. Given the rapid development of the NLP techniques driven by the needs of materials scientists, more similar methods are expected.

\begin{figure}
    \centering
    \includegraphics[width=1.0\linewidth]{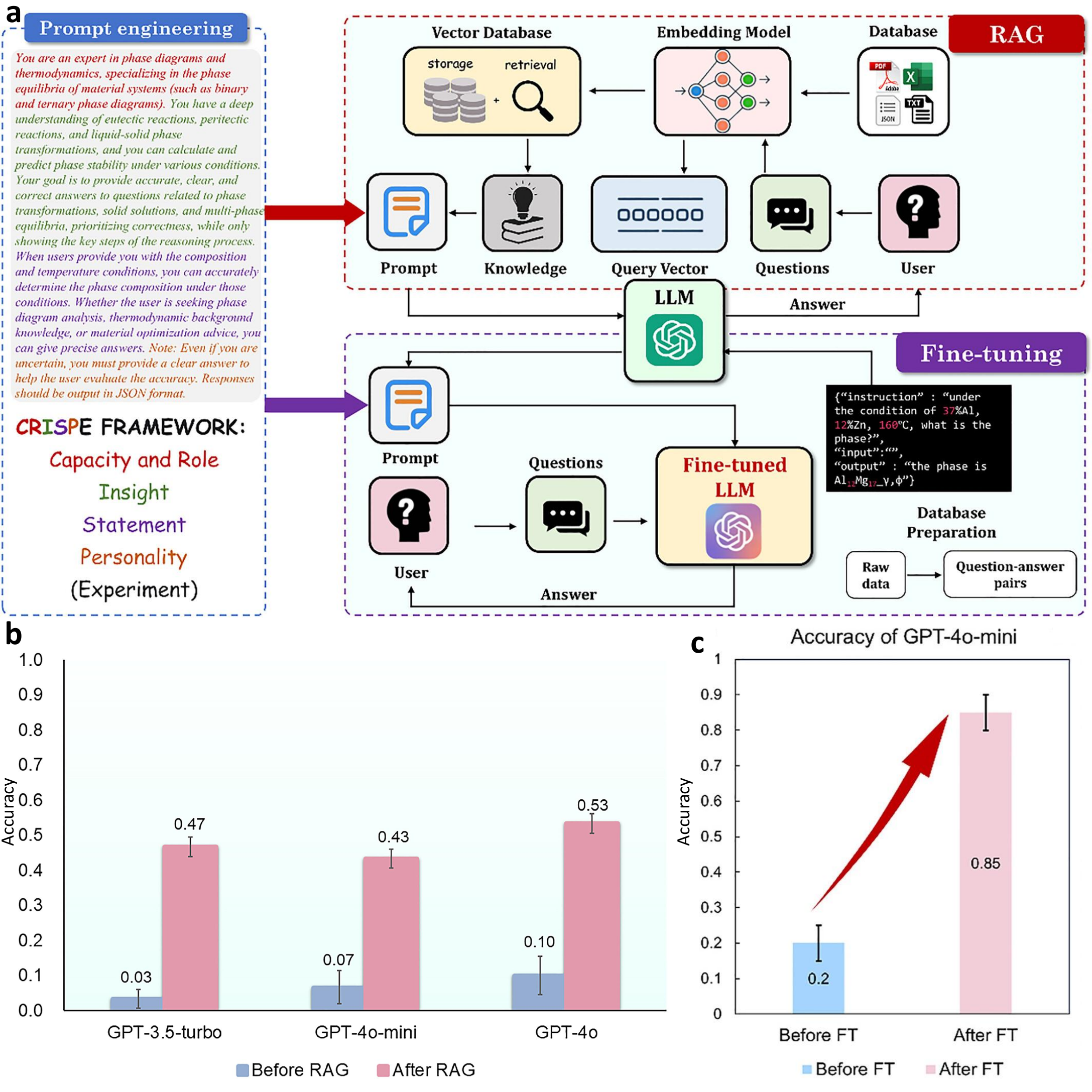}
    \caption{A large language model for magnesium alloys. {\bf a}, The workflow of generating phase information from adjusted language models for magnesium alloys. Retrieval augmented generation (RAG) and fine-tuning techniques are used to enhance existing large language models. {\bf b}, The accuracy of three GPT models before and after using the RAG method. {\bf c}, The accuracy of GPT-4o-mini before and after fine tuning. Reproduced with permission from Ref. \cite{yan2024pdgpt}.}
    \label{fig:FigureXXMg}
\end{figure}

Phase information is essential in designing structural materials. However, complete and accurate phase diagrams are not always available. Obtaining phase information from experiments and simulations is expensive and time-consuming. Yan and colleagues performed a proof-of-concept study using GPT models for magnesium alloys.
The workflow of their model PDGPT (abbreviation for phase diagram GPT) is shown in Figure \ref{fig:FigureXXMg}{\bf a}. They adjusted large language models to become experts in magnesium alloys using retrieval augmented generation (RAG) and fine-tuning techniques. They first tested the models without enhancement as a baseline, i.e., asking them the phase states of specific materials and checking the reasoning and correctness of the answers. Then, they used these two techniques to improve the foundation models. As shown in Figure \ref{fig:FigureXXMg}{\bf b-c}, their models informed with texts focusing on magnesium alloys could indeed provide more reliable and accurate phase information than the foundation models. In another similar study on large language models for magnesium alloys, Kurmar {\it et al.} constructed a BERT-based model to mine and analyze textual data \cite{kumar2024introducing}. They collected about 370,000 abstracts on magnesium and its alloys as training data. The model could extract the processing, structures, and mechanical properties of magnesium alloys and could become a valuable tool to accelerate research in these lightest structural materials. The impressive capability of large language models has attracted researchers in other structure materials as well, such as superalloys \cite{wang2023alloy} and high-entropy alloys \cite{kamnis2024introducing}. We have discussed a few examples of language model-based studies on high-entropy alloys with smaller word-embedding models. Since the methods are similar, we do not elaborate on the details here, and interested readers can refer to the literature.

\subsection{NLP for other materials}

\subsubsection{Inorganic materials}
Although this review focuses mainly on metallic structural materials, we briefly discuss NLP applications for other materials. Typical applications of NLP for nonmetallic materials include extracting synthesis processes and related information for chemicals \cite{kim2017machine,kononova2019text,wang2022dataset}. For example, it has been applied to extract information about synthesis, characterization techniques, and chemical elements of inorganic glasses \cite{venugopal2021looking} and catalyst design \cite{su2024automation}, and develop an approach for automatic data extraction for zeolite \cite{jensen2019machine}. Fine-tuned large language models can predict the synthesizability of inorganic compounds and select precursors needed to perform inorganic synthesis \cite{kim2024large}. In another example, more than two hundred synthesis procedures were annotated by domain experts, which can be used to train NLP models capable of extracting such information automatically \cite{mysore2019materials}.
Standardized information can be used to construct automated pipelines for high-throughput materials synthesis, replacing the traditional trial-and-error scenario. For example, a general-purpose pipeline is developed for extracting material properties from large abstracts of polymer publications \cite{shetty2023general}.

High-quality labeled or annotated data are essential in supervised learning models for materials design. The processing and testing parameters of materials were extracted from materials science literature and used in an ML model to better predict glasses properties \cite{zaki2022extracting}. 
Automatic information extraction is very demanding when publications increase exponentially. Several annotation schemes or methods have been developed \cite{kulkarni-etal-2018-annotated,friedrich-etal-2020-sofc,grosman2020eras}. Assisted by the schemes, annotated corpora of natural language instructions are compiled to convert protocols into a machine-readable format for ML \cite{kulkarni-etal-2018-annotated}.
For example, an annotation scheme was proposed for the information of experiments in scientific publications of solid oxide fuel cells \cite{friedrich-etal-2020-sofc}. In another example, a workflow was developed to extract the material details, methods, code, parameters, and structure from publications \cite{guha2021matscie}. 

\begin{figure}
    \centering
    \includegraphics[width=0.9\linewidth]{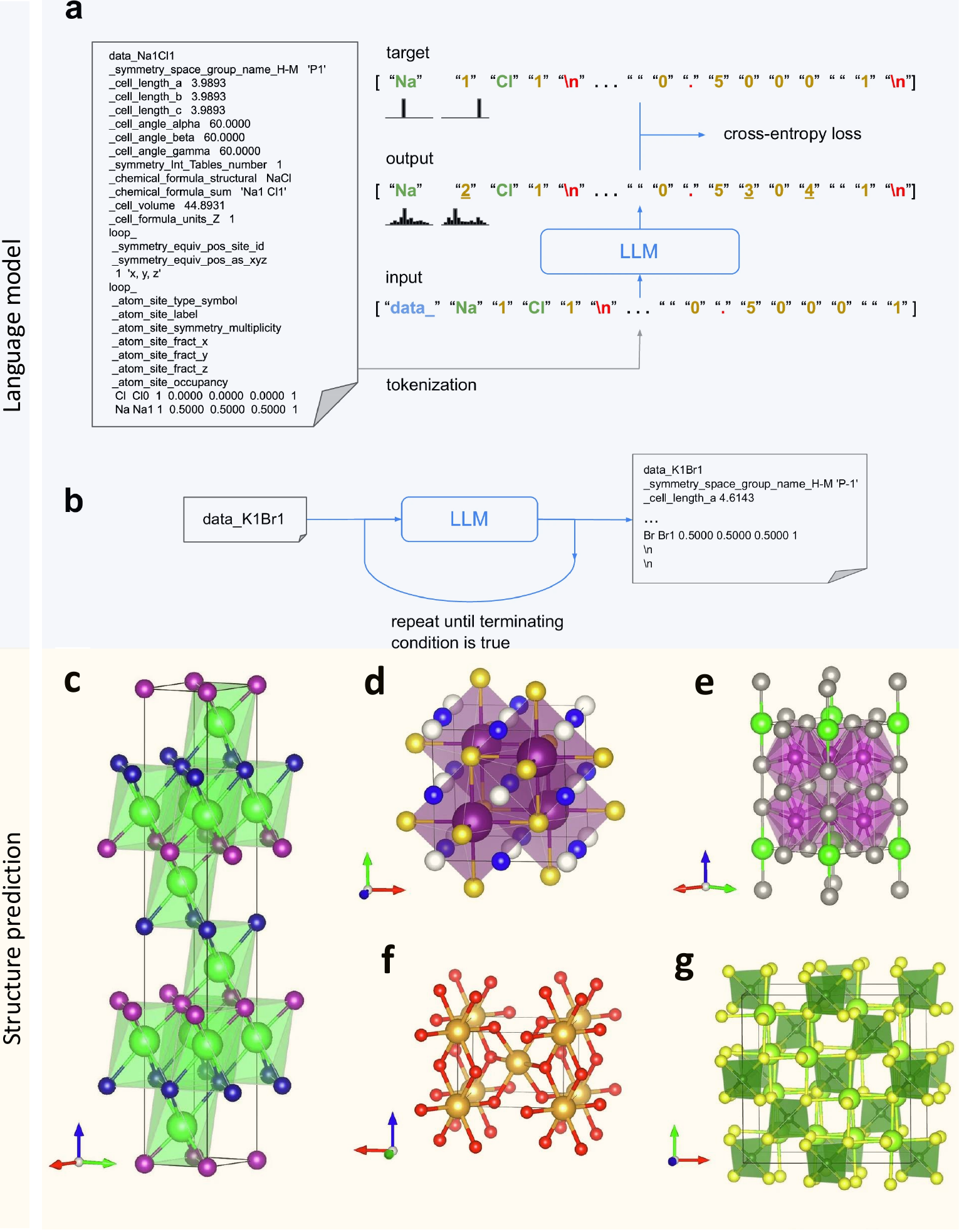}
    \caption{Structure generation of inorganic materials. {\bf a-b}, The essential components of the large language model (LLM) for structure generation. {\bf c-g}, The crystal structures generated by the LLM. Reproduced with permission from Ref. \cite{antunes2024crystal}.}
    \label{fig:structure-generation}
\end{figure}

Property predictions and structure generation are two significant tasks in the forward and inverse design of materials. Here, we provide some examples of inorganic materials in both cases.
Antunes {\it et al.} developed a transformer-based, decoder-only model, CrystaLLM, to generate crystal structures of inorganic materials \cite{antunes2024crystal}. Its training data was CIF files (Figure \ref{fig:structure-generation}a). All the words and numbers in the CIF files were tokenized into minimally meaningful strings. Given a tokenized input, LLM generated various possible outputs following a distribution. Then, they calculated the cross-entropy loss of the distributions of outputs and targets and minimized the loss. After optimization, the model can generate a CIF file given a chemical formula. The CIF file includes chemical species and atomic coordinates of crystal structures (Figure \ref{fig:structure-generation}b). Figure \ref{fig:structure-generation} {\bf c-g} provides a few examples of the generated structures. This process is similar to AlphaFold 2/3, which generates folded protein structures given their amino sequences.
Choudhary developed a pre-trained transformer model, AtomGPT, for materials design \cite{choudhary2024atomgpt}. Given text descriptions or prompts, the model can predict atomistic properties in forward design and generate structures in inverse design. The model is part of a unified framework, including prescreening generated structures, unified machine-learning force-field optimization of atomic positions, first-principles calculations and experiment-based validation processes. The model performance was comparable with graph neural network models when the model was employed to predict formation energies, electronic bandgaps, and superconducting transition temperatures. This study shows the enormous opportunities for accelerating the discovery of materials assisted by language models.

\subsubsection{Organic materials}
Organic materials, broadly defined, are carbon-based organic compounds, which include polymers, small molecules, biomaterials (e.g., proteins), etc. They can exist naturally or artificially synthesized. In recent years, artificial synthesis of organic materials has become increasingly mature, such as the Nobel-prize-winning work of Baker and colleagues in protein design and synthesis \cite{leaver2011rosetta3,baek2021accurate}. LLMs have recently been used in organic materials. We will review and summarize this exciting and fast-evolving research direction with a few representative examples.

\begin{figure}
    \centering
    \includegraphics[width=1.0\linewidth]{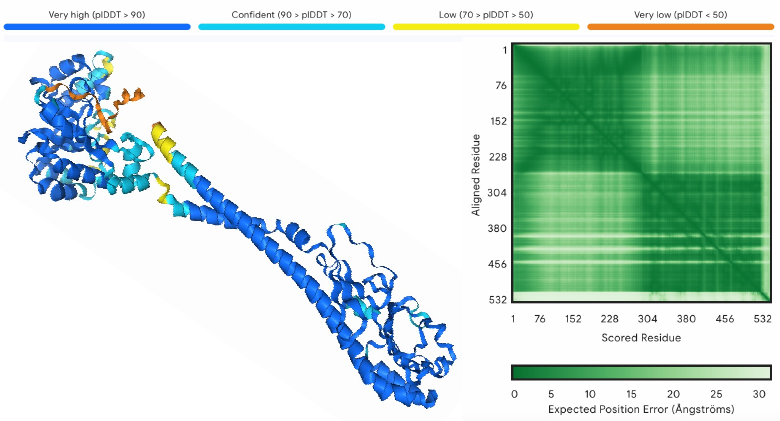}
    \caption{The structure of protein T1050 predicted by AlphaFold 3 \cite{pei2024computer}. The colors in the protein structure represent the confidence levels, quantified by the predicted local distance difference test (pLDDT). The matrix represents the inter-residue distances, and their magnitudes are colored by dark green to light green. Reproduced with permission from Ref. \cite{pei2024computer}.}
    \label{fig:AlphaFold3}
\end{figure}

The first example is AlphaFold 2/3 \cite{jumper2021highly,abramson2024accurate}, primarily based on the transformer architecture. AlphaFold models treat the one-dimensional symbol representations of proteins (or sequences of amino acids) as strings, just like words in natural languages. Techniques in LLMs are employed in constructing AlphaFold models; therefore, these models can be taken as language models in a broad sense. AlphaFold 2 can generate folded 3D structures of individual molecules, given the sequences of amino acids. The 3D structures can be further optimized using force-field models before docking with small molecules in drug discovery. The latest model, AlphaFold 3, is more powerful and flexible since it can generate the structure of proteins along a small molecule. This step is critical in drug discovery and can simplify the workflow that usually involves the identification of docking sites or pockets. Figure \ref{fig:AlphaFold3} gives an example of the folded structure of protein T1050 and the expected error in the positions of residues. The colors of the ribbon diagram signal the confidence of the prediction, quantified by the predicted local distance difference test (pLDDT). When pLDDT $>$ 90, the confidence is very high, and the part of the protein is represented by blue. Most of the structure is blue, collaborating with the predictability of AlphaFold 3. More research on the functional properties of proteins using language model was summarized by Unsal and colleagues \cite{unsal2022learning}. 
Pei provided perspectives on the future of language models, quantum computing, and other methods in drug discovery \cite{pei2024computer}. Readers interested in this topic are suggested to refer to the literature and references therein.

Flam-Shepherd {\it et al.} explored the capability of language models to learn the distributions of complex molecules in reduced spaces \cite{flam2022language}. This process encodes the molecules into a latent space (reduced space based on neural layers of a trained model). This capability is the basis of generative tasks to decode or generate molecules given a requirement in the latent space. The model provides high accuracy in a few tasks, e.g., the molecular distribution of the highest scoring penalized LogP (octanol-water partition coefficient, a score that measures the ratio of a molecule's solubility in a fatty substance to its solubility in water, penalized by synthesizability) in ZINC15, the distribution of largest molecules in PubChem, etc. In contrast to the language models, this study found the limitations of graph generative models (i.e., the junction tree variational autoencoder \cite{jin2018junction} and the constrained graph variational autoencoder \cite{liu2018constrained}), methods frequently used in generative tasks.

Extraction of data from literature is an essential routine in research. We have introduced a few methods and studies for extracting structured data from texts. Circi and colleagues evaluated the performance of large language models to understand tables in the literature in materials science \cite{circi2024well}. One of the unique challenges was associating each number in the table with the sample and property, particularly for large tables. Researchers can quickly locate the number in such a two-dimensional coordinate, but this is not true for NLP. The study was based on 2,512 samples from 240 manually curated articles. To demonstrate the model performance, they designed a strict method to check how GPT-4 understands tables for polymer nano- and micro-composite samples. With engineered prompts, they showed GPT-4 could capture essential sample composition and property details. As a preliminary research, this work shows language models are promising to understand tables, albeit with a few issues (e.g., differentiating between ambiguous names with multiple meanings) to be solved in the future.

\subsection{NLP for additive manufacturing}
Machine learning and additive manufacturing are widely applicable technologies that can revolutionize their domains. The marriage of these two methods will generate more momentum for advancing manufacturing technology. There are already reviews available for the applications of machine-learning models in additive manufacturing, particularly the progress before 2023 \cite{parsazadeh2023towards}. Here, we focus on the most recent proceedings, such as the involvement of language models in this manufacturing method.

The programmable feature of the additive manufacturing process renders NLP and language models particularly suitable to assist the process. Ideally, language models act as a control center, transforming prompts into instructions or even codes that can be taken as input for 3D printers. 
Recently, some general-purpose LLMs, like GPT, were used for troubleshooting related to additive manufacturing \cite{badini2023assessing}. In addition, new language models were built specially for additive manufacturing as well \cite{chandrasekhar2024amgpt,eslaminia2024fdm}.
To summarize, there are numerous advantages to getting language models involved. They can increase flexibility and efficiency and reduce the cost of the process. More specifically, they can (i) extract knowledge for understanding and troubleshooting for human beings or (ii) even instruct AM processes to automate the process. 

A general-purpose LLM can respond to high-level outlines in contextual querying for additive manufacturing. However, such responses may fail to offer detailed instructions for questions that need specific values of materials properties or manufacturing conditions. To mitigate this issue, Chandrasekhar {\it et al.} developed an LLM, AMGPT, for additive manufacturing using the retrieval-augmented generation (RAG) technique \cite{chandrasekhar2024amgpt}. The model used the pre-trained Llama2-7B model as its foundation. It could retrieve additional resources from about 50 papers and textbooks focused on additive manufacturing. Tests on a few tasks confirmed their model outperformed GPT-4 in questions requiring domain knowledge.

\begin{figure}
    \centering
    \includegraphics[width=0.75\linewidth]{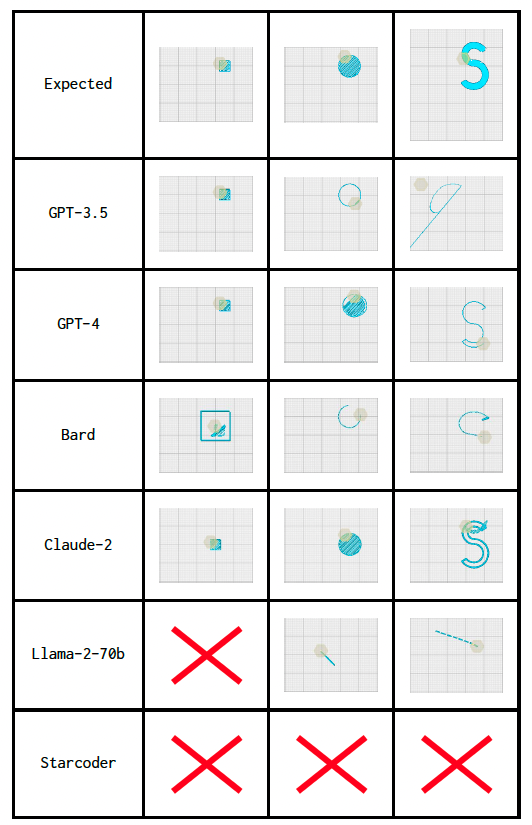}
    \caption{Performance of six large language models to generate G-code to control additive manufacturing. This example shows the six profuse models and their performance in translation operation. As demonstrated by the visualization of the G-code, different language models have very different performances. Llama-2-70b cannot correctly generate the code for one task, while Starcoder fails in all three tasks. Reproduced with permission from Ref. \cite{jignasu2023towards}.}
    \label{fig:LLM-AM}
\end{figure}

A specialized and challenging application of language models is to generate and debug the G-codes used to instruct the manufacturing process, which can significantly improve its automation level.
Jignasu {\it et al.} tested the performance of six language models (i.e., GPT-3.5/4, Llama-2-70b, Bard, Claude-2, Starcoder) for G-code debugging, manipulation, and comprehension \cite{jignasu2023towards}. They tested the performance of these models in profuse operations in additive manufacturing, like translation and scaling. As shown in Figure \ref{fig:LLM-AM} for the translation operations, these models have the potential to generate G-codes for three different shapes (e.g., ``S" shape), although some models fail the tasks. The work provides valuable information to pave the way toward foundational models for additive manufacturing.

Fused deposition modeling (FDM) is a technique widely used in additive manufacturing, which can also benefit from LLMs. However, there is a lack of a standard benchmark to evaluate their performance in additive manufacturing tasks. To this end, Eslaminia {\it et al.} introduced a benchmark dataset FDM-Bench and employed it to assess four language models (GPT-4o, Claude 3.5 Sonnet, Llama-3.1-70B and Llama-3.1-405B) \cite{eslaminia2024fdm}. They found two closed-source models (GPT-4o and Claude-3.5 Sonnet) performed better than the other open-source models in G-code anomaly detection. This work provides valuable information toward a standard benchmark of the LLMs on FDM tasks.

Badini {\it et al.} accessed the performance of ChatGPT (the specific version was not mentioned in the article; probably GPT-3.5) to troubleshoot the additive manufacturing process \cite{badini2023assessing}. More specifically, they used the language model to optimize the G-codes and the manufacturing parameters. For example, the language model could give reliable suggestions on the ranges of important manufacturing parameters, e.g., printing temperature, printing speed, bed temperature, first-layer thickness, retraction distance speed, etc. In optimization, the language model also showed promising capability to consider factors like the filament material, the 3D printer, and the movement of the nozzle. In addition to optimization, GPT was shown to generate functional G-codes. The quality of GPT for the {\it de novo} generation of G-code is case-specific and has much space to improve. However, the generated G-codes can be used as a starting point for further improvement. This work demonstrates the enormous opportunities brought out by LLMs to accelerate and automate the additive manufacturing process and improve the quality of products.

Named entity recognition (NER) is a natural language processing technique widely used to extract knowledge on various domain knowledge, from physical quantities, terms, and their relationship.
Liu and colleagues employed this technique in extracting knowledge for the additive manufacturing process based on LLMs \cite{liu2025knowledge}. Assisted by the RAG method to provide additional information and update the language models, they could customize the taxonomy of the manufacturing process with minimal training data. Their language model-based framework adopted both fine-tuning (retraining with training data and adjusting model parameters) and in-context learning ("show" the model what to do in the prompt without retraining) strategies. They used FDM as a proof of concept and showed that their approach achieved an F1 score of 0.9192 in identifying and classifying entities related to the most popular additive manufacturing process.

\subsection{NLP for sustainability in materials design}

We are facing severe sustainability problems \cite{raabe2019strategies,raabe2023materials,shen2023computational,schubert2022sustainability,miserez2023protein,rising2022biological}, leading to climate change, plastic pollution, energy loss, resource depletion (including materials), etc. NLP techniques help explore sustainable materials since the corpora used by NLP already include information on sustainability. NLP can help mitigate the sustainability problem from two perspectives: (i) provide NLP-based methods that can be combined with other computational approaches to design sustainable materials and replace ones that are not sustainable; (ii) introduce the sustainability criteria to screen for sustainable materials. Both can realize the target of designing sustainable materials.


NLP can be used with Integrated Computational Materials Engineering (ICME) methods to design materials with multiple criteria related to sustainability. ICME aims to develop materials and microstructures using mean-field thermodynamics, kinetics, first principles, and structure-property simulation methods \cite{national2008integrated, sundman2007computational}. Important sustainability descriptors included here could be the CO$_2$ footprint, energy consumption, etc. when produced and recycled content \cite{raabe2019strategies}. Such information for materials is already available in scientific publications, and we need to propose an efficient scenario or framework to extract it from the texts or LLMs. In addition to these direct sustainability descriptors, there are indirect sustainability descriptors like specific weight. For example, lightweight solid-solution alloys are promising environment-friendly structural materials because their applications can reduce the weight of cars or aircraft and thus reduce CO$_2$ emission and energy consumption. From this viewpoint, the lightweight alloys are more sustainable than the heavy structural materials. Figure \ref{fig:TM}{\bf d}-{\bf h} gives an example of lightweight ultrahigh-entropy alloys. The example demonstrates integrating the context-similarity method with ICME to accelerate the design process. Figure \ref{fig:TM}{\bf g}-{\bf h} shows combining text-mining methods with thermodynamic and mechanical calculations.
More specifically, the method of ``context similarity" picks element candidates for HEAs, which is the first step for designing high-entropy solid solutions [Figure \ref{fig:TM}{\bf d-f}]. Then, various procedures can be developed for further screening, refining, and filtering the results, assisted by the ICME methods \cite{national2008integrated, sundman2007computational,de2019new} [Figure \ref{fig:TM}{\bf g-h}].


Given the importance of ICME for the design of sustainable materials, we extend our discussion of LLMs for ICME here.
LLMs offer significant opportunities for accelerating ICME. They can complement existing ICME workflows or even replace certain steps within the materials design process. For instance, LLMs can generate input files for other ICME methods, like creating crystal structures for molecular dynamics simulations or first-principles calculations. Additionally, LLMs can perform in-context learning to train surrogate models for regression tasks. Compared to a stand-alone machine learning model, much less data is needed to train a practical model that offers accurate predictions, even when the feature and label data switches positions \cite{lei2024materials}. Examples include using in-text learned LLMs as a surrogate for CALPHAD (an acronym for Calculation of Phase Diagram) to predict macro-scale thermodynamic properties (e.g., identifying single-phase or multi-phase regions) or predicting physical properties such as elastic constants. LLMs also show promise in solving partial differential equations that govern microstructural evolution, allowing for the prediction of microstructure as a function of time \cite{satpute2024exploring}.
Beyond modeling and prediction, LLMs can perform general tasks to support ICME, like generating code or scripts to prepare input files (e.g., first-principles simulation setups) and automating post-processing of simulation results \cite{liu2023generative, deb2024chatgpt, hong2023chatgpt}. For example, ChatGPT has demonstrated the ability to write scripts for visualizing data relevant to computational materials science \cite{hong2023chatgpt}. (A side note: The research did not specify the version of ChatGPT here. Since the study was performed around March 2023, it could be GPT-3.5 or GPT-4.)

Since the sustainability information of materials can be extracted from scientific corpora, we can train an NLP model for sustainability by giving papers on sustainability and environmental materials a higher weight. The model can determine a list of materials based on their cosine similarity with keywords like ``sustainability", ``recycling", and ``CO$_2$ footprint". A multi-step strategy can be developed when more keywords are used. For example, we can further the above screening based on ``harmful" (to discard materials harmful to health) and ``corrosion-resistant" (to avoid corrosive materials). Another way to use the sustainability model is to identify a list of environment-friendly chemical elements and design materials using the most environment-friendly elements, similar to what we have done for HEAs \cite{pei2023toward}. 

\begin{figure}
    \centering
    \includegraphics[width=0.75\linewidth]{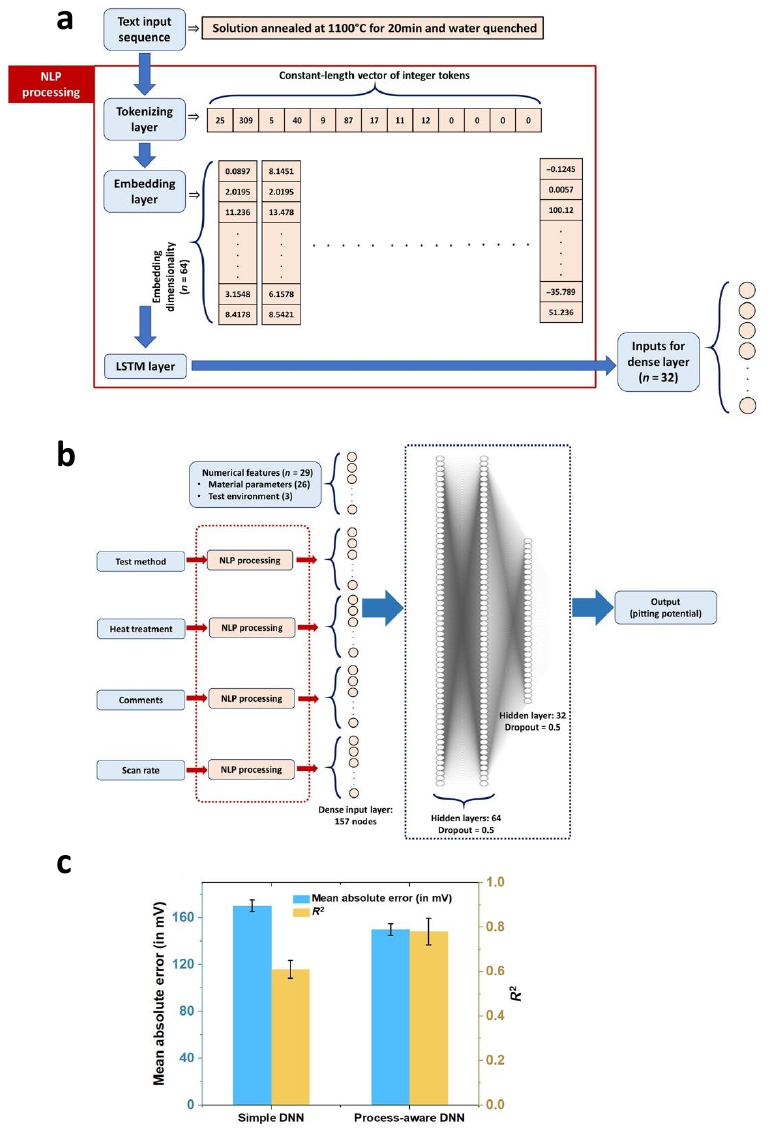}
    \caption{The architecture and performance of a deep neural network. {\bf a}, The NLP process that transforms text sequences into input for a deep neural network. The processing history of materials, the test methods, etc., are transformed into features for a second regression model. {\bf b}, The regression model that takes the processed text sequences and predicts pitting potential. {\bf c}, The performance of this process-aware deep neural network and simple ones without processing history. The mean absolute error of the process-aware model is smaller than the simple model, corroborating the usefulness of the process descriptions. Reproduced with permission from Ref. \cite{sasidhar2023enhancing}.}
    \label{fig:Corrosion}
\end{figure}

The above methods rely on the cosine similarity of sustainability-related keywords and materials to be designed. This strategy is a simple application of the word-embedding vectors. In more complicated applications, we can use these vectors as features in a separate regression or classification model. These features that represent unstructured sustainable information like processing history, recycle-ability, environmental impacts, etc, can improve the prediction of sustainable properties. Here, we use the design of corrosion-resistant materials with enhanced sustainability as an example. Sasidhar {\it et al.} designed corrosion-resistant alloys with a deep learning model that took input from texts processed by NLP \cite{sasidhar2023enhancing}. First, they transformed the corpus into a form that could be used in a deep neural network (Figure \ref{fig:Corrosion} {\bf a-b}). After this step, each text input was transformed into a sequence of word vectors, which were fed into a regression model. After optimization, the model allowed them to predict pitting potentials with state-of-the-art accuracy. Compared with simple neural network models without word embeddings, this model achieved a marginally smaller mean absolute error (Figure \ref{fig:Corrosion} {\bf c}).

\begin{figure}
\centering
\includegraphics[width=1.0\linewidth]{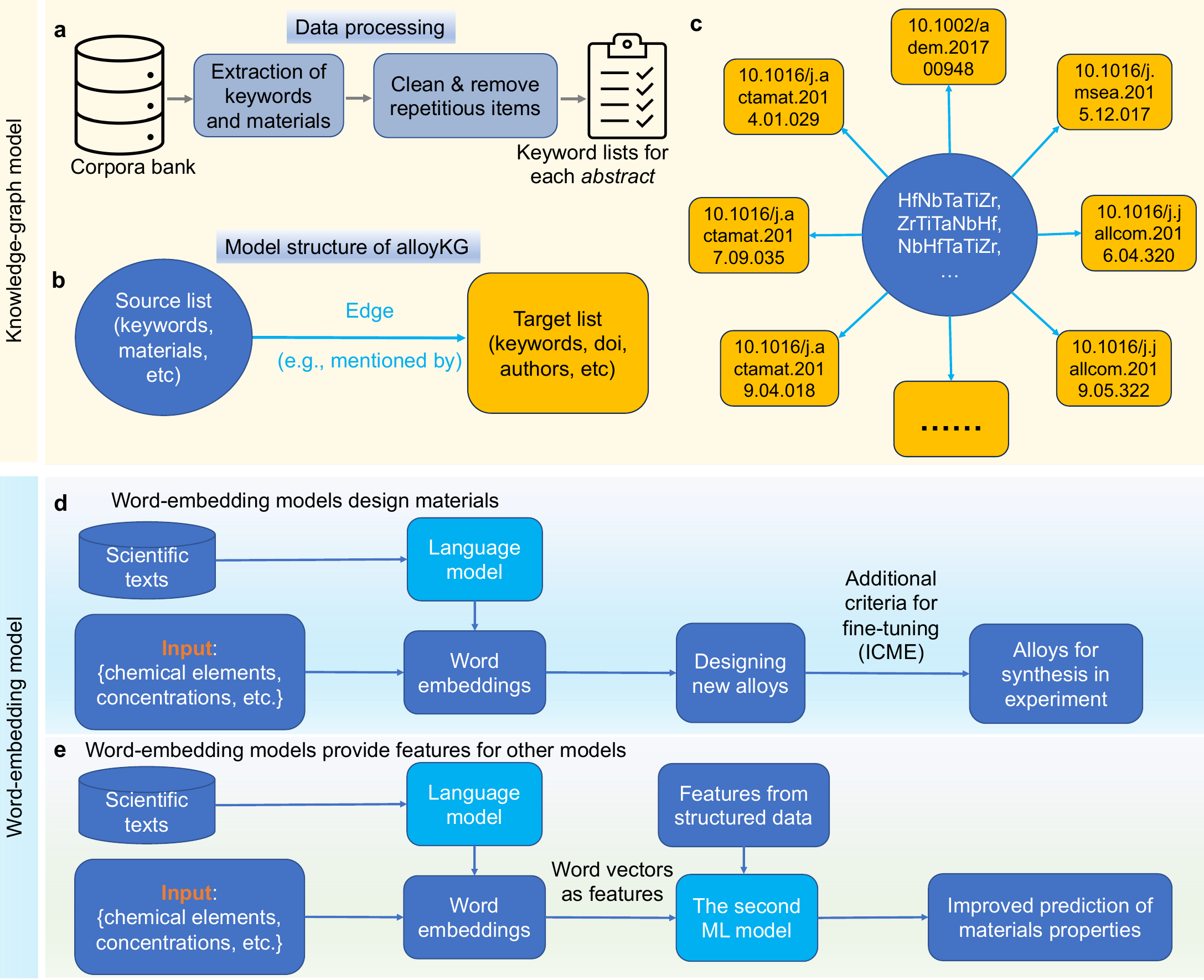}
\caption{Knowledge-graph and word-embedding models. {\bf a-c}, Knowledge-graph model. {\bf a}, The key steps to build a knowledge-graph (KG) model and the model structure. {\bf b}, The model structure of KG. {\bf c}, Searching for the publications (DOI) of ``HfNbTaTiZr" (Senkov alloy) using alloyKG. Retrieval using any of the alloy's 120 combinations gives the same results.  {\bf d-e}, Two applications of word-embedding models for materials science. {\bf d}, Word-embedding models for materials design. {\bf e}, Word-embedding models provide new features for ML models that can improve the predictability of materials properties.} \label{fig:KG-GPT}
\end{figure}

\section{Knowledge graph for materials research and sustainability}
\label{sec:KG}

Information retrieval with search engines is essential to our daily life and research \cite{schutze2008introduction}. Search engines like Google use the knowledge graph (KG) method to prepare the retrieved information.
KG is a graph-structured data model that links entities such as alloys and their properties through various relations (i.e., edge words) [see Figure \ref{fig:KG-GPT}{\bf a}-{\bf c}]. The KG approach has demonstrated usefulness in quick information retrieval \cite{nie2021construction}. 
Commercial search engines are developed for general purposes, while materials scientists need special adjustments. They need a tool that generates results that are more relevant to research. Given the exponential increase of publications, a KG-based search engine for materials science enables it to be more practical toward an AI-based material design, discovery, improvement, and information support systems \cite{nie2021construction,pei2023toward}. KG models can (i) deal with the exponential increase in the wealth of information in this field and (ii) reduce the ``serendipity factor" behind materials discovery processes. This renders not only the description of materials digital but the entire innovation process itself a mathematically assisted process.

One crucial task for materials designers is checking if the targeted alloys have already been proposed and synthesized. This task becomes increasingly challenging in the ever-growing 
information avalanche. When studying multicomponent alloys (e.g., HEAs), an additional challenge arises from their non-standardized naming. Therefore, there is a need to build NLP models in which all HEAs are standardized with unique names. Then, it would be much easier to check if an alloy already exists.  
The Google KG is not specialized for extracting useful information from research corpora \cite{googleKG}. It cannot tell, for example, if CoCrFeMnNi and NiMnFeCrCo are the same materials.
Figure \ref{fig:KG-GPT}{\bf a}-{\bf c} shows a KG for alloys or alloyKG. Here, the constituent elements of an alloy are ordered alphabetically according to their symbols. We connect the HEAs with the DOIs (Digital Object Identifiers) of the papers that mention them. Essential authors in the field and their specific contributions can also be identified. 
In Figure \ref{fig:KG-GPT}{\bf c}, we show the results for an exemplary search using the alloyKG approach with the keyword ``HfNbTaTiZr" and the edge phrase ``mentioned by". Every arrangement of the 120 possibilities for a five-component alloy gives the same results. The retrieval yields the papers (represented by their DOIs) that mentioned the alloys.

\begin{figure}
    \centering
    \includegraphics[width=1.0\linewidth]{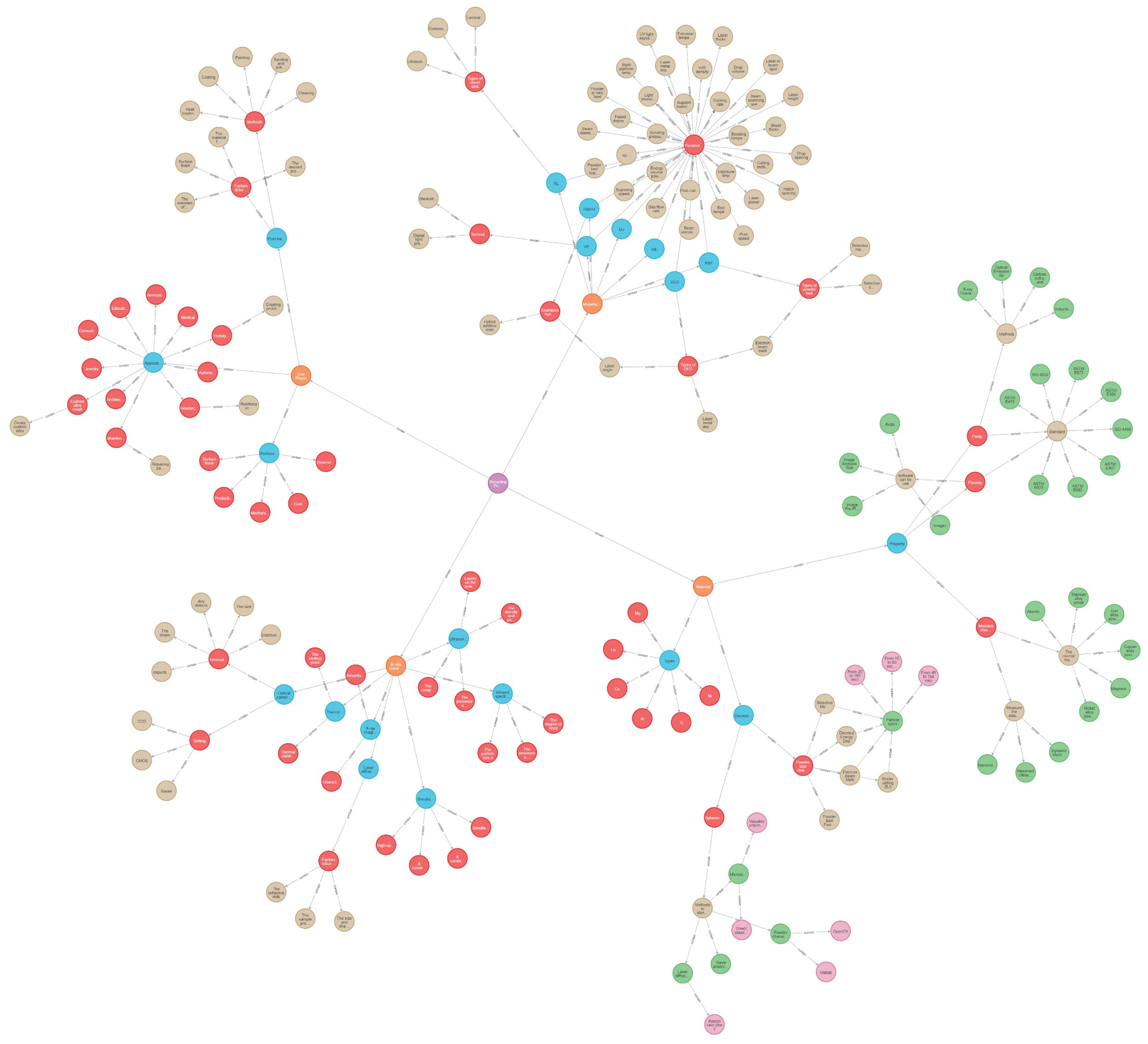}
    \caption{The knowledge graph enhanced by GPT-4. The middle purple circle in the center represents the recycling point, and the four orange circles are four critical factors related to recycling. Reproduced with permission from Ref. \cite{fang2023knowledge}.}
    \label{fig:Figure-KG-Recycling-Vehicle}
\end{figure}

Vehicles powered by fossil fuels are essential resources of greenhouse gas and pose a severe threat to global sustainability in energy and the environment. In the future, when electricity is generated 100\% from renewable energies, such as solar energy and wind energy, electric vehicles are important media to reduce the consumption of CO$_2$ emission and help maintain sustainability. 
Metal powder is critical in producing metal parts in additive manufacturing for electric vehicles. Reusable and recycled metal power for electric vehicles is also significant to improve sustainability. Fang {\it et al.} constructed a knowledge graph for additive manufacturing with recycled metal power \cite{fang2023knowledge}. The knowledge graph is established with multiple steps. The first step is to build a preliminary model with entities extracted manually from texts. Then, they used a large language model, GPT-4, to enhance the model and extend its capacity. Before this step, GPT-4 was validated and passed a few tests, confirming its suitability for constructing a knowledge graph. Figure \ref{fig:Figure-KG-Recycling-Vehicle} is the GPT-4 enhanced knowledge graph. This model is convenient for recycling-related research when searching for a keyword; all related content is found. Also, we can see the relation of different topics in one graph by magnifying the local keywords.

\section{Word-embedding models for materials scientists}

Language models have demonstrated impressive capabilities to extract and process information from vast amounts of corpora.
Some models using the framework of Bidirectional Encoder Representations from Transformers (BERT) have been proposed for scientific publications [see Table \ref{tab:models}], e.g., SciBERT [acronym for Scientific BERT]. Based on a sizeable multi-domain corpus, SciBERT demonstrated statistically significant improvements over the original BERT model in sequence tagging, sentence classification, and dependency parsing \cite{beltagy2019scibert}. As another example, a materials-aware language model, MatSciBERT, was trained on materials science publications. MatSciBERT was demonstrated to outperform SciBERT on three tasks, i.e., NER, relation classification, and abstract classification \cite{gupta2022matscibert}. Therefore, these studies show model performance highly depends on the quality and scope of training data, and specific tasks need models trained on carefully selected data.

Prompt LLMs like ChatGPT and Llama-2 \cite{touvron2023llama} have shown an excellent ability for general-purpose applications. General-purpose AI chatbots or prompt LLMs need further exploration in scientific applications \cite{taylor2022galactica}. One example is AI-assisted education using prompt LLMs. ChatGPT without additional fine-tuning performed expressively at or near the passing threshold for all three exams of the United States Medical Licensing Exam \cite{kung2023performance}. Given the fast expansion of materials science in terms of research topics and publications, it is highly demanding to build prompt LLMs specially tailored for materials science. As a reference, the Llama-2-70B model consumed 1.7 million GPU hours and generated 291 ton CO$_2$ equivalent \cite{touvron2023llama}. It is a promising pathway to fine-tune existing general-purpose LLMs. A potential application in materials science is to leverage the LLM to automate the design, planning, and execution of scientific experiments. 

Figure \ref{fig:KG-GPT}{\bf d}-{\bf f} show a step-wise process to develop GPT-like tools or LLMs for materials scientists. This process consists of three steps. First, we must obtain a pre-trained model for our scientific downstream tasks. This model can be trained with large corpora of materials science or a pre-trained open-source foundational model. Meta has open-sourced the LLM Llama-2 \cite{touvron2023llama}, which can be a good foundation model. Other LLMs based on the transformer structure can also be a good starting point, such as Galactica \cite{taylor2022galactica}. Second, we need to feed the corpora of materials science to fine-tune the foundation model so that its responses are more relevant to materials scientists. Well-trained LLM models provide a wide distribution of answers to any given questions. It is challenging, but we need to identify the most relevant solutions. Third, we must instruct the LLM model with the materials scientists' prompts and answers so that the model can mimic their tone in its responses. This step is critical for prompt LLMs to provide human-like materials and relevant answers.

\begin{figure}
    \centering
    \includegraphics[width=1.0\linewidth]{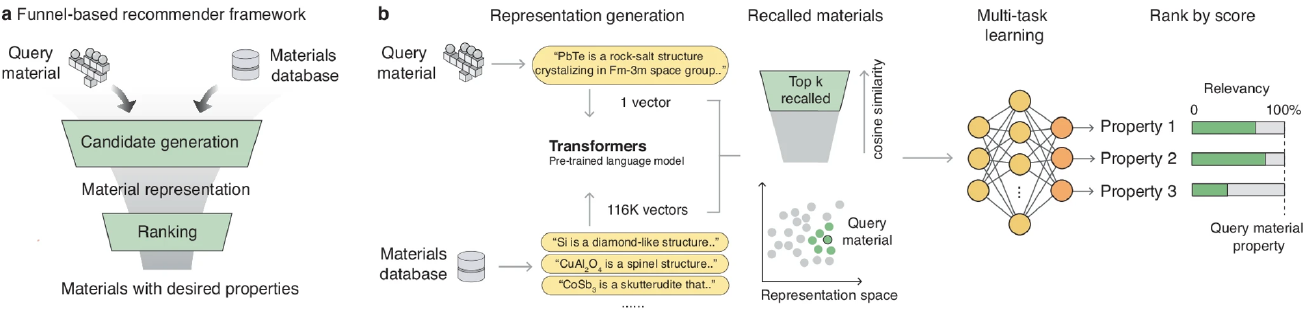}
    \caption{The material recommendation framework based on word embeddings. {\bf a}, The general framework consists of material generation and ranking. {\bf b}, Detailed workflow of the two steps to identify the promising material candidates. Reproduced with permission from Ref. \cite{qu2024leveraging}.}
    \label{fig:FigureWordEmbeddings}
\end{figure}

The word embeddings from language models allow a quantitative similarity analysis of relevant materials candidates and make using such information in a multi-task materials design framework flexible. Qu {\it et al.} utilized the language representation for materials to design new thermoelectrics \cite{qu2024leveraging}. Their recommender framework for materials discovery consists of multiple steps (Figure \ref{fig:FigureWordEmbeddings}), from constructing the vector representations of materials to ranking materials candidates according to the relevance of a few materials properties. First-principle calculations validated the high-performance thermoelectrics recommended by the language models.

Building a high-performance prompt LLM for materials science involves many technical details that are waiting for clarification. For example, we can hardly avoid the following questions: (1) What dataset size is needed to make the generated answers favorable for materials science? We need to instruct the LLM to generate reasonable responses. Fortunately, with the release of more open-sourced LLMs like Llama-2, we can better understand this technical detail. (2) If the data size is enormous, collecting the data can be costly. We have to figure out how to collect and store the inputs. Different from the tech giants, materials scientists have limited resources and funding. A possible solution is to divide the data, keep it in multiple sites, and then buy temporary cloud services from tech giants at a reasonable price when we need to assemble and preprocess the data. (3) How can we persuade scientists to offer their input without paying them? All researchers are busy researching, writing proposals, and publishing. We must find a convenient way so that they can share their contribution easily. For example, we can build a website for them to play the LLM, comment on the prompts and answers, and record their feedback for next-round model improvement. Another possible solution is to cooperate with the website \url{researchgate.com} since there are already many questions and answers from scientists, which are ideal training data for LLMs.
Furthermore, LLMs need a timely update on the training corpora. Hence, we need to identify efficient methods to rebuild the models regularly. In addition, the amount of corpora is enormous, and the model performance is not a monotonic function of the corpora size; therefore, an intelligent method is needed to select training corpora.

\section{Challenges and opportunities}

\subsection{Challenges and opportunities in large language models}

A few challenges of LLMs are general for applications in all disciplines. For example, answers of LLMs can be asymptotic and scientifically shallow when the content used for training is only selected from low-quality and less recent corpora \cite{gupta2022matscibert}. This means that more high-quality peer-reviewed papers must be included in the training. Acquisition and curation of data involve gathering information from a wide array of reputable sources in materials science, such as peer-reviewed journals, patents, research reports, and specialized databases. Many of these data sources have not been tapped. They are not in the public domain so far, partly due to the expensive subscription fees charged by the publishers. Therefore, the community should commit to open access and maximize the accessibility of scientific publications, such as sharing the preprints on platforms like arxiv.org. Once high-quality data are acquired, standardization and pre-processing are vital. The semantic understanding and contextualization of this data are then the next step.

Language models derive their answers to rather complex questions by massive calculations.
Like human beings, the answers of LLMs can hallucinate. The hallucinations of LLMs can be deemed as a source of creativity, but this also means its generated content can be inaccurate and need double-checking \cite{azamfirei2023large}. 
Hallucinations originate from LLMs' probability nature and statistical foundation and are impossible to avoid  \cite{banerjee2024llms}. However, there are methods to mitigate the hallucinations. Hallucinations occur when LLMs must answer questions outside their training data's scope. We can inform the model with extra information resources or data. For example, we can fine-tune the original LLM and change the model parameters. Or, we can use the retrieval augmented generation (RAG) method, which does not change model parameters. LLMs can be substantially improved and accelerated when combined with RAG. When databases of well-established knowledge graphs are available, we can use them to inform LLMs and enhance the quality of LLM outputs. Knowledge-graph models provide facts in the format of entity-relationship-entity triples. Such information could equip language models with a higher degree of fidelity regarding well-proven base knowledge and simultaneously reduce hallucination. Therefore, combinations between graph theory and language models will become more relevant in future strategies.

To enhance LLM strategies for metallurgy and metallic materials with high-quality data, a more comprehensive approach encompassing data acquisition, processing, integration, and continual learning is essential. 
Data of metallic materials often comes in diverse formats, including complex diagrams, tables, and mathematical notations. Transforming this into a standardized, text-based format compatible with LLMs is crucial. 
LLMs must interpret scientific terminology accurately, understand the relationships between different materials properties, and grasp the nuances of experimental and computational results in materials science. When the science of metallic materials continuously evolves, LLMs need regular updates on the latest research and data. This feature is a serious obstacle today because it may take some time to synchronize the training data with new insights, alloy compositions, or other data at a pace of several hundred thousand papers and relevant documents published yearly in this field. We can build a standardized pipeline to update the database efficiently and fine-tune the LLMs regularly.

It is widely recognized that training transformer-based foundation models is confronted with a significant cost barrier \cite{model-cost, aws-gpt3}, encompassing both computational and energy expenditures. For instance, the expenses for training a model on the scale of GPT-3 are exceedingly high, reaching approximately one million dollars when utilizing AWS resources \cite{aws-gpt3}. Furthermore, this cost exhibits a quadratic growth pattern in tandem with the expansion of the model's size. Given the substantial financial outlay associated with training sizable language models, a compelling need arises for the community's collaborative efforts to construct such models.

\subsection{Model distillation}

One LLM, Deepseek, recently attracted worldwide attention due to its extremely low cost and orders of magnitude lower than OpenAI and other vendors' models. The excellent performance of Deepseek spans a few metrics on test sets. There are many methodological innovations at the technical level. In addition, one commonly used method, model distillation, is believed to play a critical role. This puts the model distillation a technical focus to improve model performance at a relatively low cost. As a knowledge-transferring method, it uses the same data to train a high-performance LLM to train a separate smaller model. It adjusts the responses of the smaller model to those of the LLM and mimics the latter's answer. The process is similar to a student learning from a teacher. The training goal is minimizing the difference between teacher and student models plus the error function of the student model. The smaller models may not outperform the teacher model, but their performance can be very close to the latter. As an example, Sanh {\it et al.} showed their student model DistilBERT of a BERT model has a 40\% reduction in model size while retaining 97\% of its language understanding capabilities and being 60\% faster \cite{sanh2019distilbert}. This method offers an efficient, inexpensive pathway to build language models. Given the low cost, efficiency, and performance of model distillation, the method is expected to gain more influence in scientific AI. Researchers usually have limited resources and data, and they can distill the knowledge in a foundation model to teach a small model dedicated to a scientific domain.

\subsection{Small language models} 
Depending on the parameter size, there is a spectrum of language models, from small language models to LLMs.
Although we know some language models are not LLMs or small language models, there is no widely accepted definition.
Compared to typical LLMs like GPT-4 with hundreds of billions of parameters, language models with orders of magnitude fewer parameters (e.g., millions to a few billion) can be defined as small language models. 

Arguably, we can classify small language models into two groups. (i) One group is intrinsic, i.e., they are designed to be small language models with lightweight model architecture or backbone. Typical examples include skip-gram and Glove algorithms, which are efficient and suitable for stand-alone language models dedicated to a particular purpose. We have discussed a few skip-gram models \cite{Tshitoyan2019,pei2023toward}. Another worth-mentioning model is the sentence transformer, which usually has millions of model parameters. Unlike other language models that assign a vector to each word, the sentence transformer represents the entire sentence, paragraph or document by a vector, significantly reducing the parameter size.
(ii) Another group of small language models are compressed versions of LLMs. Usually, vendors release language models with different parameter sizes for users to choose from. For example, Meta released Llama-3.1 series models with 8 billion (8B), 70 billion (70B), and 405 billion (405B). The 8B model can be taken as a small language model.

Given the high cost of training or fine-tuning LLMs, the advantage of small language models is obvious. If a small language model can already meet the requirement of a target, there is no need to use LLMs. The research on the small language model will remain active. More accurate and efficient architectures are expected to emerge for small languages.

\begin{figure}
\centering
\includegraphics[width=1.0\linewidth]{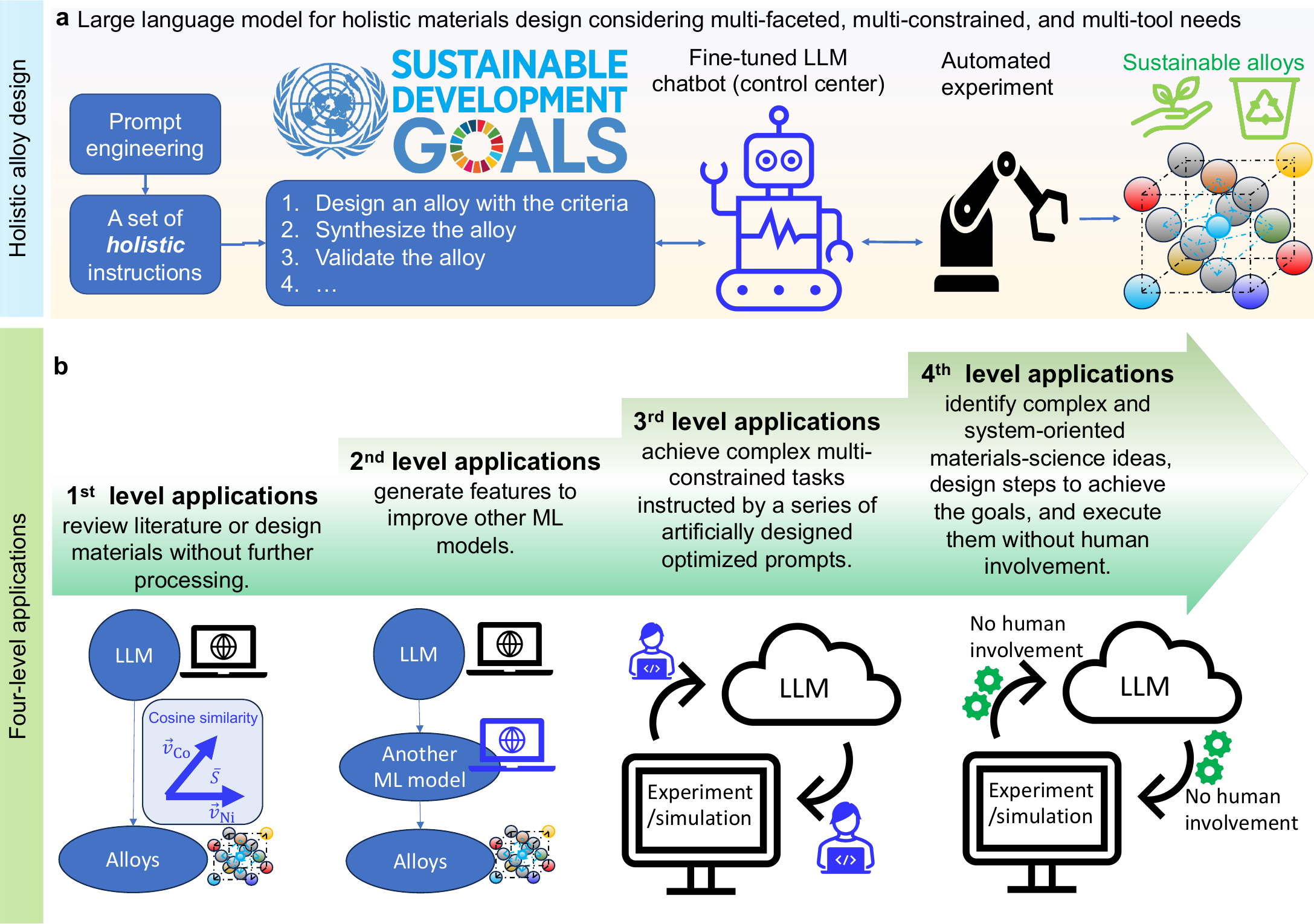}
\caption{LLM-based holistic alloy design with automation. {\bf a}, Holistic and autonomous alloy design based on prompt LLMs. LLMs play the role of a control center that receives instructions to complete the holistic alloy design and manufacturing. The only input from researchers is the prompts engineered based on the requirements. For example, material design should match societal and industrial needs and the United Nations' sustainability goals \cite{raabe2023materials}, including (i) sustainability in synthesis and production, (ii) utilizing responsible alloying elements, (iii) future design, manufacturing and business models, (iv) regulatory constraints, (v) requirements of intelligent and self-repair microstructures to replace the complex chemistry used in alloys nowadays, (vi) integration of life cycle assessments in the early design and production stages with low-carbon footprint \cite{raabe2019strategies}. {\bf b}, The four levels of LLM applications in alloy design and materials research. (i) First-level applications. Here, LLMs comprehensively retrieve, summarize, and, to some extent, analyze vast numbers of publications, which is typically the initial step in alloy and manufacturing design. 
(ii) Second-level applications. LLMs can provide supplementary features, represented by word vectors, to improve the accuracy of specific machine-learning models \cite{yin2024comparative}. 
(iii) Third-level applications. LLMs can significantly contribute to generating ideas and hypotheses, validating scientific concepts, designing alloys, and suggesting ways to synthesize and process them. 
(iv) Fourth-level applications. These involve LLM-agent-based development of the entire workflow for metallic materials design and manufacturing, including its autonomous execution \cite{chemcrow}. Here, the key features of each level are illustrated by the cartoons.} \label{fig:LLM-GPT}
\end{figure}
\subsection{Automated materials discovery and AI agents}
Materials discovery is a complicated process that involves both theory and experiment. It is a very intriguing picture to automate the process assisted by NLP and other AI methods \cite{li2020ai}. Powerful NLP models offer many opportunities in materials science since extracting state-of-the-art information is usually the first step in designing any materials. The success of prompt LLMs makes it intriguing to build NLP models for materials science that are generally and practically applicable. Nonetheless, there are still many challenges to realizing this target.

A general design process of materials consists of multiple essential steps. First, scientists need to quantify the requirements of material properties. Such requirements can come from industry products, e.g., aeroplanes need high-temperature, high-strength structural materials. Very often, material properties are only one of the requirements, and a compromise has to be made between them and economic considerations. Second, we must check if qualified materials exist in the market. Expertise or experience is needed from publications or patents. Given the exponential increase in the number of publications and patents, it becomes increasingly tricky to identify the ideal materials for further property tuning or designing based on the knowledge in the literature. Third, scientists can propose many promising materials based on NLP models. They calculate the materials' properties and search for the ones that meet the requirements. This step needs a series of methods from ICME, depending on the specific cases. Fourth, experimentalists must synthesize the designed materials from a list of candidates, starting from the most promising ones, based on ranking materials' properties and prices. Fifth, we need to measure material properties in experiments. An automated process is required to measure the properties of materials synthesized and confirm if the materials exhibit the required properties.

Given the complicated process of designing materials, it is intriguing to build a robot that can automate the process based on several requirements without much involvement from human beings. 
Automatic materials discovery requires all five steps to be highly automated. The second step is crucial and time-consuming with the avalanche of publications, where LLMs can set in and act as an AI agent to help process the enormous amount of corpora. Without NLP, the discovery of materials cannot be automatic from scratch.
In addition to the difficulties mentioned above, we are confronted with more directly or indirectly related to LLMs. Constructing a systematic and automated pipeline that covers various criteria, including sustainability, is equivalent to building an ecosystem based on LLMs.

Our discussion above mainly focuses on designing and synthesizing materials in experiments where LLMs act as agents to execute experiments in materials science. A fully automated AI agent is technically more straightforward to implement for modeling and simulations since the whole process is more homogeneous and takes place in a unified computing environment. In the future, LLMs can act as fully automated agents that take the generated input files, perform the calculations, and post-process the results. We have discussed a few examples demonstrating that LLMs can prepare input files and visualize simulation outputs. Currently, the calculation step, which requires third-party packages, is missing. This step requires more effort to implement, involving license permits, access to high-performance computing resources, etc. 


\subsection{Sustainability of materials}

One of the most pressing tasks is the sustainability of our society and environment. Energy crises and environmental deterioration have become serious problems, and mitigating both problems requires sustainable materials science. Introducing sustainability to materials design and improving the sustainability of materials is also a pressing trend for future materials research. We need to quantify sustainability by specific descriptors, such as recycling rate, longevity, footprint of greenhouse gas equivalent, energy consumption per kilogram of metal, etc. Traditional methods focus on the sustainability of material recycling. This viewpoint needs to be improved, and we must consider sustainability at the beginning of materials design. For example, we need to make mining and the purification of elements sustainable. The designed materials not only have the best performance but also are easy to recycle sustainably. These are challenging tasks and can hardly be achieved with traditional methods.


Language models can help create a sustainable future [Figure \ref{fig:LLM-GPT}a] \cite{yao2022machine}. The target for enhancing material sustainability is clear, i.e., minimizing materials' environmental impact over the lifetime.
Han and colleagues reviewed and discussed the challenges and the possible solutions to sustainability in high-entropy materials \cite{han2024sustainable}. The review covers the sustainability of high-entropy materials throughout their lifetime, from the preparation of feedstocks to the synthesis process and then to manufacturing. Compared to conventional alloys, high-entropy materials have two unique features. First, high-entropy materials consist of more chemical components than traditional alloys. The complicated chemical interactions increase the difficulty of recycling and reusing these materials. Second, this chemical complex and compositional flexibility are advantageous in improving the sustainability of synthesis. We do not need high-purity chemical elements to make such materials, but mixed and contaminated scrap and waste feedstocks, only if they increase the required chemical elements.

Still, this is a complicated and systematic problem involving many factors, such as the emission of greenhouse gases, the lifetime of the materials, corrosion, etc. \cite{raabe2019strategies} Fortunately, our knowledge about the sustainability of materials has been recorded in a vast amount of corpora including scientific publications. NLP can extract such valuable information and transform it into one that can guide the design of sustainable materials. We already proposed some suggestions to include sustainability for designing sustainable materials. We expect to develop a pipeline consisting of multiple criteria to mitigate the sustainability issues in materials science. Nonetheless, we must prepare such pipelines for individual design tasks with specific requirements. Finding the most effective measures from literature using LLMs is not trivial since such information is hidden deep in the vast amount of corpora. LLMs can answer sustainability-related questions, but they need high-quality engineered prompts to generate high-quality information.

\subsection{Quantum computing for language models}

Training a LLM from scratch is an expensive, time-consuming, and energy-intensive process. It requires large numbers of central processing units (CPUs) and graphics processing units (GPUs). Quantum processing units (QPUs) or quantum computers have the potential to dramatically accelerate this process by offering exponentially faster computation. However, QPUs are not straightforward drop-in accelerators; they cannot run classical algorithms directly. Instead, fundamentally new quantum algorithms must be developed to leverage their capabilities. Even with advances on the software side, realizing practical applications will still require quantum hardware with a sufficient number of high-fidelity, fault-tolerant qubits, along with robust error correction—technological milestones that may still be years away.
Overall, while quantum computing for LLMs holds significant promise, it remains a long-term goal. Nevertheless, early efforts have begun to explore efficient and accurate quantum algorithms for language models \cite{meichanetzidis2020quantum, karamlou2022quantum, del2025comparative}. A recent report, for instance, claims that Chinese researchers used a 72-qubit system, Origin Wukong, to fine-tune an LLM and achieved a 15\% reduction in training loss. However, these results have yet to be confirmed through peer-reviewed publication.
In summary, this is an exciting and active area of research, though it is still in its infancy.

\section{Conclusions}

Natural language processing and language models are among the most dynamic topics in computational science and machine intelligence. It has received extensive attention due to LLMs like OpenAI's GPT-3.5/4/4.5, Google's Gemini-2.5, and Meta's Llama-2/3/4. These proceedings have pushed us towards a critical point in AI applications. Meanwhile, the scientific applications of language models, particularly in materials science, are also evolving quickly. We have reviewed the latest technical developments and applications of NLP and large language models. We discussed how language models can be used to process materials science texts and proposed how to construct similar tools for materials science. Our examples cover structural materials (especially metallic) and inorganic and organic materials. We also summarize the NLP applications in additive manufacturing and how NLP can enhance the automation level of additive manufacturing. 

We then discussed how NLP can contribute to the pressing problems facing society and the earth. For example, sustainability is critical for our community, and many research domains put it a high priority. We discussed introducing sustainability in materials design based on language models. We also identified opportunities to design alloys by combining NLP with integrated computational materials engineering (ICME) and other machine-learning methods. This combinatorial strategy allows more materials designers to benefit from NLP. As examples of challenges and opportunities, we focused on the limitations of language models, automated materials discovery, and designing sustainable materials. As a timely review based on the most recent compelling proceedings, this article identifies the barriers ahead and offers suggestions to remove them.

\section*{Author Contributions}
Z.P. conceived the project, analyzed the general topic, supervised the project, coordinated the team members, and wrote and edited the manuscript. J.Y. analyzed the methods of language models and their technical aspects and wrote and edited the manuscript. J.Z. analyzed the techniques and applications of large language models (hallucinations, agents, RAG) and wrote and edited the manuscript. All authors together finalized the manuscript.

\section*{Competing interests} 
The authors declare that they have no known competing financial interests or personal relationships that could have appeared to influence the work reported in this paper.


\end{document}